\documentclass[aps,prd,longbibliography,reprint,twocolumn,amsmath,amssymb,amsfonts,showpacs,footnote,superscriptaddress]{revtex4-2}

\usepackage[T1]{fontenc}
\usepackage{lmodern}
\usepackage{amsmath,amssymb,amsfonts,mathrsfs}
\usepackage{bm}
\usepackage{mathtools}
\usepackage{physics}
\usepackage{hyperref}
\usepackage{xcolor}
\usepackage{booktabs}
\usepackage{siunitx}
\usepackage{tabularx}
\usepackage{url}

\usepackage{orcidlink}

\definecolor{navyblue}{rgb}{0.0, 0.0, 0.5}
\definecolor{ferrarired}{rgb}{1.0, 0.11, 0.0}
\definecolor{persianblue}{rgb}{0.11, 0.22, 0.73}
\usepackage{float}

\hypersetup{
colorlinks=true,
citecolor=blue,%
linkcolor=ferrarired,%
urlcolor=persianblue,%
filecolor=blue,
linktoc=page
}

\newcommand{\ii}{\mathrm{i}}

\providecommand{\ii}{\mathrm{i}}

\newcommand{\draftfigure}[2][0.98\linewidth]{%
\IfFileExists{#2}{\includegraphics[width=#1]{#2}}{%
	\fbox{\parbox[c][4.2cm][c]{0.92\linewidth}{\centering
			\textbf{Figure placeholder}\\[0.4em]
			\texttt{\detokenize{#2}}}}%
}%
}

\begin{document}

\title{Exceptional points and the breaking of Schwarzschild isospectrality under local potential perturbations}

	\author{Mohamed \surname{Ould~El~Hadj}\,\orcidlink{0000-0002-8558-7992}}
\email{med.ouldelhadj@gmail.com}
\affiliation{No-affiliation}

\date{\today}

\begin{abstract}
The odd- and even-parity gravitational perturbations of a Schwarzschild black hole are governed by the distinct Regge-Wheeler and Zerilli potentials, yet their quasinormal-mode (QNM) spectra are identical in vacuum. Building on our previous study of the odd-parity Regge-Wheeler spectrum [M. Ould El Hadj and S. R. Dolan, arXiv:2608.16283], we investigate the corresponding even-parity Zerilli problem and compare the two sectors under the same perturbation of amplitude $\epsilon$ at radius $r_0$. Using the Chandrasekhar-Detweiler relation, we derive an exact relation between the normalized Wronskians and show that Schwarzschild QNM isospectrality is not preserved under an identical nonzero localized defect. We find that the even- and odd-parity repelling-point branches approach the same Schwarzschild QNM frequencies at large $r_0$, with their first parity-dependent separation appearing at order $1/\mathcal R^2$, where $\mathcal R=r_{0*}/M$ is the dimensionless tortoise-coordinate location of the defect. The corresponding exceptional points (EPs) form two shifted sequences with the same asymptotic spacing in the tortoise coordinate, while their relative displacement is fixed by the phase of the asymptotic Chandrasekhar-Detweiler factor. The critical perturbation amplitudes share the same exponential decay rate in the two sectors, whereas the ratio of their asymptotic prefactors is fixed by the modulus of the same factor. We further find that the even- and odd-parity EPs connect the same neighboring QNM overtones, with nearly identical arclengths for the corresponding spectral bridges. Finally, the exceptional degeneracy is parity selective: at identical values of $(r_0,\epsilon)$, a pair of QNMs may coalesce at an EP in one parity sector while remaining distinct in the other, leading to a corresponding difference in the local pole structure of the ringdown response.
\end{abstract}

\maketitle

\tableofcontents

\section{Introduction}
\label{sec:Introduction}

The direct detection of gravitational waves from binary black hole coalescences has made black hole ringdown accessible to observational study~\cite{LIGOScientific:2016aoc,Carullo:2019flw,Isi:2019aib}. In black hole perturbation theory, ringdown is characterized by quasinormal modes (QNMs), a discrete set of complex resonances whose real and imaginary parts determine the oscillation and damping timescales of the response~\cite{Echeverria:1989hg,Leaver:1985ax,Leaver:1986gd,Berti:2009kk}. The QNM spectrum thus provides a characteristic spectral fingerprint of the black hole and of the dynamics governing its perturbations, forming the basis of black hole spectroscopy and offering a way to test gravity in the strong-field regime~\cite{Dreyer:2003bv,Berti:2005ys,Berti:2018vdi,Berti:2025hly}.

The use of QNM spectra as probes of black holes naturally raises the question of their stability under small perturbations. Black holes are open dissipative systems: perturbations can escape to spatial infinity or cross the event horizon, and the corresponding QNM boundary-value problem is intrinsically non-Hermitian. As recognized long ago by Nollert~\cite{Nollert:1996rf}, small changes in the effective potential can produce unexpectedly large displacements of quasinormal frequencies. This spectral sensitivity has subsequently been explored from several complementary viewpoints, including environmental perturbations, pseudospectral analysis, and localized potential deformations~\cite{Barausse:2014tra,Jaramillo:2020tuu,Jaramillo:2021tmt, Cheung:2021bol,Destounis:2021lum,Torres:2026uey}. The effect is particularly pronounced for QNM overtones, whose frequencies may undergo large migrations under perturbations that are small at the level of the effective potential.

Exceptional points (EPs) provide a natural framework for understanding the nonlinear structure underlying such spectral migrations. In a non-Hermitian spectral problem, an EP is a parameter value at which two or more eigenvalues and their associated states coalesce, producing a branch-point singularity and a multi-sheeted spectral structure ~\cite{Ding:2022juv}. Near a second-order EP, the frequency splitting has the characteristic square-root dependence on parameter detuning, while continuation around the degeneracy exchanges the two local spectral sheets. Exceptional-point phenomena have recently emerged in a variety of black hole settings, including exact coalescences, avoided crossings, resonant excitation, hysteresis, and exceptional lines ~\cite{Motohashi:2024fwt,Cavalcante:2024swt,Cavalcante:2024kmy, Yang:2025dbn,Oshita:2025ibu,PanossoMacedo:2025xnf, Cao:2025afs,Nakamoto:2026lyo,Cavalcante:2026vgr}. These developments have shown that EPs are not merely isolated degeneracies, but can organize the migration, connectivity, and excitation properties of neighboring QNM branches. Related recent work has developed a resolvent-based description of scattering and sourced response at black hole exceptional points~\cite{Morikawa:2026fzn}.

Within this broader setting, in Ref.~\cite{OuldElHadj:2026vym} we investigated the odd-parity Schwarzschild spectrum under a localized delta-function perturbation of the Regge-Wheeler potential. The point-defect model admits an exact resonance condition in terms of a normalized Wronskian and reveals continuous repelling-point (RP) branches together with discrete families of real exceptional points. Each family accumulates on a Schwarzschild QNM as the defect is moved outward. At large distance, the exceptional points become asymptotically equally spaced in the tortoise coordinate, whereas their critical couplings decrease exponentially according to a law governed by the complex frequency of the corresponding unperturbed QNM. Global continuation further showed that neighboring Schwarzschild overtones are connected through EP-mediated spectral bridges and that the local double-pole structure controls the characteristic response near coalescence.

Schwarzschild gravitational perturbations have an additional structure that was not involved in that analysis. They separate into odd- and even-parity sectors governed by the Regge-Wheeler and Zerilli equations ~\cite{Regge:1957td,Zerilli:1970se}. Although the corresponding effective potentials are different, their vacuum QNM spectra coincide exactly. This odd-even isospectrality is one of the classical structural properties of Schwarzschild perturbation theory and follows from the Chandrasekhar-Detweiler relation between the homogeneous solutions~\cite{Chandrasekhar:1975zza,Chandrasekhar1983}. In vacuum general relativity, parity isospectrality also persists perturbatively beyond the Schwarzschild limit in a slow-rotation expansion~\cite{Franchini:2023xhd}.

The robustness of gravitational isospectrality under deformations has recently attracted renewed attention. Parity splitting of QNM spectra is known to arise in a variety of non-Einsteinian settings, including black holes in nonlinear electrodynamics ~\cite{Chaverra:2016ttw} and higher-curvature extensions of general relativity~\cite{Silva:2024ffz,Silva:2026jih}. More generally, modifications of the gravitational perturbation equations typically lift the degeneracy between odd- and even-parity spectra, although special higher-derivative theories can preserve isospectrality, at least in the eikonal regime ~\cite{Cano:2024wzo,Bah:2026aia}. Recent work has related this special preservation to a gravitational electric-magnetic duality at the light ring~\cite{Bah:2026aia}. At the same time, the loss of frequency isospectrality need not translate in a simple way into mode excitation or observable ringdown waveforms ~\cite{Silva:2024ffz,Silva:2026jih}.

The present work brings these two lines of investigation together. Rather than modifying the underlying gravitational theory, we apply the same prescribed localized defect to the Regge-Wheeler and Zerilli potentials. The unperturbed system is therefore exactly isospectral, the deformation is identical in the two parity sectors, and any resulting spectral difference can be traced directly to the way in which the two Schwarzschild master problems respond to the same local perturbation.

The central question is therefore not simply whether the even-parity Zerilli problem exhibits the same EP phenomenology as its odd-parity counterpart. Instead, we ask which elements of the perturbed spectral organization are inherited from the common Schwarzschild QNM spectrum, and which reveal the breaking of the Chandrasekhar-Detweiler isospectral relation under the common localized perturbation. In particular, there is no a priori reason for the odd- and even-parity EPs to occur at the same frequencies, defect locations, or critical couplings once the perturbation is switched on. Nor is it obvious whether a loss of frequency isospectrality should alter the global connectivity of the QNM branches.

The point-defect model makes this comparison unusually transparent. For each parity sector $p\in\{e,o\}$, where $e$ and $o$ denote even and odd parity, respectively, we denote by $F^{(p)}(\omega;r_0)$ the normalized Wronskian that determines the exact point-defect resonance condition. The perturbed resonances are then the zeros of the spectral function
\[
D^{(p)}(\omega,\epsilon;r_0) = F^{(p)}(\omega;r_0)-\epsilon = 0.
\]
The Chandrasekhar-Detweiler relation between the homogeneous solutions induces an exact relation between the two normalized Wronskians. We show that
\[
F^{(e)}(\omega;r_0) = C(\omega;r_0)F^{(o)}(\omega;r_0),
\]
where $C(\omega;r_0)$ is the local Chandrasekhar-Detweiler factor. Identical localized perturbations thus generically break Schwarzschild QNM isospectrality, even though the vacuum spectra coincide exactly.

The breaking is nevertheless highly structured. At large defect radius, the odd- and even-parity RP branches have the same leading accumulation toward the common Schwarzschild QNM, while their difference first appears at subleading order. The corresponding EPs form two asymptotically shifted sequences with the same spacing and a finite parity offset. Their critical couplings share the same exponential fragility exponent but differ in their asymptotic prefactors. Remarkably, these two parity effects are controlled by the phase and modulus, respectively, of the same asymptotic Chandrasekhar-Detweiler factor.

The global organization of the spectrum reveals a further important aspect of the parity structure. Although the EP frequencies, radii, and critical couplings depend on parity, continuation through the first five EPs of both the $n=0$ and $n=1$ families connects the same neighboring Schwarzschild overtones in the two sectors. The corresponding spectral bridges closely track one another in the complex-frequency plane and possess nearly identical integrated arclength diagnostics. At identical control parameters, however, the exceptional degeneracy itself is parity selective: the controls that produce an EP in one sector generically leave the opposite sector nondegenerate. These results reveal a distinction between the parity-sensitive \emph{spectral placement} of exceptional points and the much more parity-robust \emph{spectral architecture} that they organize.

The paper is organized as follows. In Sec.~\ref{sec:formulation}, we review the odd- and even-parity gravitational perturbations of the Schwarzschild black hole, recall the Chandrasekhar-Detweiler isospectral relation in vacuum, and formulate the common localized point-defect problem, including the definitions of repelling points and exceptional points. In Sec.~\ref{sec:IsospectralityBreaking}, we derive the exact relation between the odd- and even-parity normalized Wronskians and show how an identical local defect breaks Schwarzschild isospectrality, both exactly and in the small-coupling regime. In Sec.~\ref{sec:LargeDistanceParityStructure}, we develop the large-distance parity asymptotics of the repelling-point and exceptional-point families, including the parity-dependent lattice offset and critical-coupling prefactors. In Sec.~\ref{sec:Results}, we present the numerical results, examining the small-coupling parity splitting, the repelling-point and exceptional-point sequences, their large-distance behavior, and the spectral connectivity of the two parity sectors. Finally, Sec.~\ref{sec:Discussion} summarizes our results and discusses possible extensions.

Throughout this article, we use geometrized units $G=c=1$ and assume a harmonic time dependence of the form $e^{-\ii\omega t}$ for the perturbative fields.

\section{Schwarzschild perturbations with localized defects}
\label{sec:formulation}

\subsection{Odd- and even-parity perturbations of the Schwarzschild black hole}
\label{subsec:SchwarzschildPerturbations}

We briefly recall the main ingredients of the gravitational perturbation problem of the Schwarzschild black hole and establish the notation and conventions that will be used throughout this article.

The exterior of a Schwarzschild black hole of mass $M$ is described by the metric
\begin{equation}
	\dd s^2 = -f(r)\,\dd t^2 +f(r)^{-1}\,\dd r^2 +r^2\dd\sigma_2^2,
	\label{eq:SchwarzschildMetric}
\end{equation}
where
$
	f(r)=1-\frac{2M}{r},
$
and
$
	\dd\sigma_2^2 = \dd\theta^2+\sin^2\theta\,\dd\varphi^2
$
denotes the line element on the unit two-sphere \(S^2\). The Schwarzschild coordinates satisfy \(t\in(-\infty,+\infty)\), \(r\in(2M,+\infty)\), \(\theta\in[0,\pi]\), and \(\varphi\in[0,2\pi]\).

We shall also use the tortoise coordinate \(r_*\in(-\infty,+\infty)\), defined in terms of the Schwarzschild radial coordinate by
\begin{equation}
	\frac{\dd r}{\dd r_*}=f(r).
	\label{eq:TortoiseDerivative}
\end{equation}
It is explicitly given by
\begin{equation}
	r_*(r) = r+2M\ln\left(\frac{r}{2M}-1\right)+r_*^{(0)},
	\label{eq:TortoiseCoordinate}
\end{equation}
where \(r_*^{(0)}\) is an arbitrary additive constant. Unless otherwise specified, we set $r_*^{(0)}=0$. The function \(r_*=r_*(r)\) provides a bijection from \(r\in(2M,+\infty)\) to \(r_*\in(-\infty,+\infty)\). Hence, the event horizon and spatial infinity correspond, respectively, to \(r_*\to-\infty\) and \(r_*\to+\infty\).

After separation of variables, the radial functions $\phi_{\omega\ell}^{(e)}(r)$ and $\phi_{\omega\ell}^{(o)}(r)$ satisfy, respectively, the Zerilli and Regge-Wheeler equations~\cite{Regge:1957td,Zerilli:1970se}
\begin{equation}
	\left[ \frac{\dd^2}{\dd r_*^2} +\omega^2 -V_\ell^{(e/o)}(r) \right] \phi_{\omega\ell}^{(e/o)}(r) =0.
	\label{eq:ZM_RW}
\end{equation}
Here and in the following, the superscripts $(e)$ and $(o)$ refer, respectively, to even (polar) and odd (axial) quantities, according to their parity under the antipodal transformation on the unit two-sphere $S^2$.

The effective potentials appearing in Eq.~\eqref{eq:ZM_RW} are the Zerilli potential
\begin{align}
	V_\ell^{(e)}(r) ={}& \frac{f(r)} {r^3(\Lambda r+6M)^2} \Big[ \Lambda^2(\Lambda+2)r^3 +6\Lambda^2Mr^2 \notag\\ &\hspace{6em} +36\Lambda M^2r +72M^3 \Big],
	\label{eq:Ve}
\end{align}
and the Regge-Wheeler potential
\begin{equation}
	V_\ell^{(o)}(r) = f(r) \left[ \frac{\Lambda+2}{r^2} -\frac{6M}{r^3} \right].
	\label{eq:Vo}
\end{equation}
In Eqs.~\eqref{eq:Ve} and \eqref{eq:Vo}, we have introduced
\begin{equation}
	\Lambda=(\ell-1)(\ell+2)=\ell(\ell+1)-2.
	\label{eq:LambdaDefinition}
\end{equation}

For each parity sector, we introduce two linearly independent solutions of Eq.~\eqref{eq:ZM_RW}, denoted by $\phi_{\omega\ell}^{\mathrm{in},(e/o)}(r)$ and $\phi_{\omega\ell}^{\mathrm{up},(e/o)}(r)$.

The functions $\phi_{\omega\ell}^{\mathrm{in},(e/o)}(r)$ are defined by their purely ingoing behavior at the event horizon,
\begin{equation}
	\phi_{\omega\ell}^{\mathrm{in},(e/o)}(r) \underset{r_*\to-\infty}{\sim} e^{-\ii\omega r_*},
	\label{eq:phi_in_horizon}
\end{equation}
while, at spatial infinity, they have the asymptotic behavior
\begin{equation}
	\phi_{\omega\ell}^{\mathrm{in},(e/o)}(r) \underset{r_*\to+\infty}{\sim} A_\ell^{(-,e/o)}(\omega)e^{-\ii\omega r_*} + A_\ell^{(+,e/o)}(\omega)e^{+\ii\omega r_*}.
	\label{eq:phi_in_infinity}
\end{equation}

Similarly, the functions $\phi_{\omega\ell}^{\mathrm{up},(e/o)}(r)$ are defined by their purely outgoing behavior at spatial infinity,
\begin{equation}
	\phi_{\omega\ell}^{\mathrm{up},(e/o)}(r) \underset{r_*\to+\infty}{\sim} e^{+\ii\omega r_*},
	\label{eq:phi_up_infinity}
\end{equation}
and, near the event horizon, they behave as
\begin{equation}
	\phi_{\omega\ell}^{\mathrm{up},(e/o)}(r) \underset{r_*\to-\infty}{\sim} B_\ell^{(-,e/o)}(\omega)e^{-\ii\omega r_*} + B_\ell^{(+,e/o)}(\omega)e^{+\ii\omega r_*}.
	\label{eq:phi_up_horizon}
\end{equation}
Here, $A_\ell^{(\pm,e/o)}(\omega)$ and $B_\ell^{(\pm,e/o)}(\omega)$ are complex amplitudes.

The Wronskian of the two solutions,
\begin{equation}
	W_\ell^{(e/o)}(\omega) = \phi_{\omega\ell}^{\mathrm{in},(e/o)} \biggl[\overrightarrow{\frac{\dd}{\dd r_*}} - \overleftarrow{\frac{\dd}{\dd r_*}}\biggr] \phi_{\omega\ell}^{\mathrm{up},(e/o)},
	\label{eq:Wronskian}
\end{equation}
is independent of $r_*$. By evaluating it at the event horizon and at spatial infinity, one obtains
\begin{equation}
	W_\ell^{(e/o)}(\omega) = 2\ii\omega A_\ell^{(-,e/o)}(\omega) = 2\ii\omega B_\ell^{(+,e/o)}(\omega).
	\label{eq:WronskianAmplitudes}
\end{equation}

We recall that the Schwarzschild QNMs are solutions which are purely ingoing at the event horizon and purely outgoing at spatial infinity. They occur when $\phi_{\omega\ell}^{\mathrm{in},(e/o)}$ and $\phi_{\omega\ell}^{\mathrm{up},(e/o)}$ become linearly dependent, i.e., when their Wronskian vanishes,
\begin{equation}
	W_\ell^{(e/o)}(\omega)=0.
	\label{eq:QNMconditionW}
\end{equation}
Equivalently, their complex frequencies are the zeros of the incoming amplitude at spatial infinity,
\begin{equation}
	A_\ell^{(-,e/o)}(\omega)=0.
	\label{eq:QNMconditionA}
\end{equation}
For the harmonic time dependence $e^{-\ii\omega t}$ adopted here, the damped Schwarzschild QNMs lie in the lower half of the complex $\omega$ plane. Moreover, the spectrum is symmetric with respect to the imaginary axis: if $\omega$ is a QNM frequency, then $-\omega^*$ is also a QNM frequency. In the following, we restrict ourselves to the frequencies with $\operatorname{Re}\omega>0$ and denote them by
\begin{equation}
	\omega_Q^{(n,e/o)},
	\qquad n=0,1,2,\ldots,
	\label{eq:QNMfrequencies}
\end{equation}
where $n$ is the overtone index and $n=0$ denotes the fundamental mode.

\subsection{Chandrasekhar-Detweiler relation and Schwarzschild isospectrality}
\label{subsec:CDIsospectrality}

Although the Zerilli and Regge-Wheeler equations involve different effective potentials, their homogeneous solutions are related by the Chandrasekhar-Detweiler transformation
~\cite{Chandrasekhar:1975zza,Chandrasekhar1983},
\begin{equation}
	(\Omega-\ii\omega)\phi_{\omega\ell}^{(e)}(r) = \left[ \frac{\dd}{\dd r_*}+\Gamma(r) \right] \phi_{\omega\ell}^{(o)}(r),
	\label{eq:CD}
\end{equation}
where
\begin{equation}
	\Omega=\frac{\Lambda(\Lambda+2)}{12M},
	\label{eq:Omega}
\end{equation}
and
\begin{equation}
	\Gamma(r) = \Omega+ \frac{6Mf(r)}{r(\Lambda r+6M)}.
	\label{eq:Gamma}
\end{equation}
The relations below are understood away from the algebraically special frequencies $\omega=\pm\ii\Omega$. The low-lying QNMs considered in the present work are well separated from these values.

Applied to the boundary-normalized solutions introduced above, the Chandrasekhar-Detweiler relation implies that the scattering amplitudes satisfy
\begin{subequations}
	\label{eq:CD_amplitudes}
	\begin{align}
		A_\ell^{(-,e)}(\omega) &= A_\ell^{(-,o)}(\omega), \label{eq:CD_Aminus} \\ A_\ell^{(+,e)}(\omega) &= \frac{\Omega+\ii\omega} {\Omega-\ii\omega} A_\ell^{(+,o)}(\omega).
		\label{eq:CD_Aplus}
	\end{align}
\end{subequations}
In particular, Eq.~\eqref{eq:CD_Aminus}, together with Eq.~\eqref{eq:WronskianAmplitudes}, gives
\begin{equation}
	W_\ell^{(e)}(\omega) = W_\ell^{(o)}(\omega).
	\label{eq:WronskianIsospectrality}
\end{equation}
The incoming amplitudes therefore possess the same zeros in the two parity sectors, and the Schwarzschild QNM spectra are isospectral,
\begin{equation}
	\omega_Q^{(n,e)} = \omega_Q^{(n,o)} \equiv \omega_Q^{(n)}, \qquad n=0,1,2,\ldots .
	\label{eq:QNMIsospectrality}
\end{equation}

\subsection{Localized perturbations of the effective potentials}
\label{subsec:LocalizedPerturbations}

We follow here the construction developed in Sec.~II.B of our previous work~\cite{OuldElHadj:2026vym}, where the resonance condition was derived for a general localized perturbation and then specialized to a delta-function defect. Rather than repeating the full derivation, we summarize below the essential steps needed for the present analysis. We refer the reader to Ref.~\cite{OuldElHadj:2026vym} for further details.

We introduce the same localized defect into the Zerilli and Regge-Wheeler effective potentials,
\begin{equation}
	V_\ell^{(e/o)}(r) \longrightarrow V_\ell^{(e/o)}(r) + \epsilon\,\delta(r_*-r_{0*}),
	\label{eq:DeltaPotentialPerturbation}
\end{equation}
where $r_{0*}=r_*(r_0)$ denotes the location of the perturbation and $\epsilon\in\mathbb{R}$ its strength. Throughout this article, the same values of $r_0$ and $\epsilon$ are used in the two parity sectors.

The corresponding radial equation is
\begin{equation}
	\left[ \frac{\dd^2}{\dd r_*^2} +\omega^2 -V_\ell^{(e/o)}(r) -\epsilon\,\delta(r_*-r_{0*}) \right] \phi_{\omega\ell}^{(\epsilon,e/o)}(r_*) =0.
	\label{eq:PerturbedRadialEquation}
\end{equation}
The perturbed radial function is continuous at $r_*=r_{0*}$, whereas its first derivative satisfies the jump condition
\begin{equation}
	\left. \frac{\dd\phi_{\omega\ell}^{(\epsilon,e/o)}}{\dd r_*} \right|_{r_{0*}^{+}} - \left. \frac{\dd\phi_{\omega\ell}^{(\epsilon,e/o)}}{\dd r_*} \right|_{r_{0*}^{-}} = \epsilon\, \phi_{\omega\ell}^{(\epsilon,e/o)}(r_{0*}).
	\label{eq:DeltaJumpCondition}
\end{equation}

A perturbed resonance is purely ingoing at the event horizon and purely outgoing at spatial infinity. Its radial function can therefore be written as
\begin{equation}
	\phi_{\omega\ell}^{(\epsilon,e/o)}(r_*)
	=
	\begin{cases}
		\mathcal N_{\rm in}^{(e/o)}
		\phi_{\omega\ell}^{\mathrm{in},(e/o)}(r_*),
		& r_*<r_{0*},
		\\[2mm]
		\mathcal N_{\rm up}^{(e/o)}
		\phi_{\omega\ell}^{\mathrm{up},(e/o)}(r_*),
		& r_*>r_{0*}.
	\end{cases}
	\label{eq:PiecewisePerturbedResonance}
\end{equation}
Using the continuity and jump conditions at $r_{0*}$ gives the exact resonance condition
\begin{equation}
	\frac{ W_\ell^{(e/o)}(\omega) }{ \phi_{\omega\ell}^{\mathrm{in},(e/o)}(r_{0*}) \phi_{\omega\ell}^{\mathrm{up},(e/o)}(r_{0*}) } -\epsilon =0.
	\label{eq:ExactDeltaSpectralCondition}
\end{equation}

It is therefore natural to introduce the normalized Wronskians
\begin{equation}
	F^{(e/o)}(\omega;r_0) = \frac{ W_\ell^{(e/o)}(\omega) }{ \phi_{\omega\ell}^{\mathrm{in},(e/o)}(r_{0*}) \phi_{\omega\ell}^{\mathrm{up},(e/o)}(r_{0*}) },
	\label{eq:NormalizedWronskians}
\end{equation}
which, using Eq.~\eqref{eq:WronskianAmplitudes}, can also be written as
\begin{equation}
	F^{(e/o)}(\omega;r_0) = \frac{ 2\ii\omega A_\ell^{(-,e/o)}(\omega) }{ \phi_{\omega\ell}^{\mathrm{in},(e/o)}(r_{0*}) \phi_{\omega\ell}^{\mathrm{up},(e/o)}(r_{0*}) }.
	\label{eq:NormalizedWronskiansAmplitude}
\end{equation}
The perturbed spectral equations take the compact form
\begin{equation} D^{(e/o)}(\omega,\epsilon;r_0) \equiv F^{(e/o)}(\omega;r_0)-\epsilon =0.
	\label{eq:PerturbedSpectralEquation}
\end{equation}

The functions $F^{(e/o)}$ are invariant under independent multiplicative rescalings of the corresponding ingoing and upgoing solutions. In the unperturbed limit, $\epsilon=0$, Eq.~\eqref{eq:PerturbedSpectralEquation} reduces to the Schwarzschild QNM condition, provided that the product of the two homogeneous solutions does not vanish at $r_{0*}$.

At fixed $r_0$, the perturbed frequencies trace trajectories in the complex-frequency plane as $\epsilon$ is varied. Differentiating Eq.~\eqref{eq:PerturbedSpectralEquation} gives
\begin{equation}
	 \frac{\dd\omega}{\dd\epsilon} = \frac{1} {F_\omega^{(e/o)}(\omega;r_0)}, \qquad F_\omega^{(e/o)} \equiv \frac{\partial F^{(e/o)}}{\partial\omega}.
	\label{eq:SpectralFlowEquation}
\end{equation}
Thus, in both parity sectors, the exact delta-perturbed spectral problem is completely determined by the corresponding normalized Wronskian $F^{(e/o)}(\omega;r_0)$.

\subsection{Repelling points and exceptional points}
\label{subsec:RepellingExceptionalPoints}

We follow here Sec.~II.C of Ref.~\cite{OuldElHadj:2026vym}, where the relation between repelling points of the spectral flow and second-order exceptional points was established in detail. We recall only the main elements needed for the present analysis of the two parity sectors.

For a fixed perturbation location $r_0$, a repelling point is defined as a stationary point of the normalized Wronskian with respect to the complex frequency,
\begin{equation}
	F_\omega^{(e/o)} \left(\omega_r^{(e/o)};r_0\right) =0,
	\label{eq:RepellingPointCondition}
\end{equation}
where $F_\omega^{(e/o)}=\partial_\omega F^{(e/o)}$. According to Eq.~\eqref{eq:SpectralFlowEquation}, a simple zero of $F_\omega^{(e/o)}$ corresponds to a pole of the spectral-flow field, and resonance trajectories passing nearby are therefore strongly deflected.

For a generic RP, $F^{(e/o)}(\omega_r^{(e/o)};r_0)$ is complex. A real EP is obtained when this value lies on the real $F$ axis, i.e.,
\begin{equation}
	\operatorname{Im} F^{(e/o)} \left(\omega_r^{(e/o)};r_0\right) =0,
	\label{eq:EPRealityCondition}
\end{equation}
with the corresponding critical perturbation strength given by
\begin{equation}
	\epsilon_{\rm EP}^{(e/o)} = \operatorname{Re} F^{(e/o)} \left(\omega_r^{(e/o)};r_0\right).
	\label{eq:EPCriticalCoupling}
\end{equation}

The exceptional nature of such a point follows directly from the exact perturbed Wronskian, which can be written as
\begin{equation}
	\mathcal W_\ell^{(e/o)}(\omega,\epsilon;r_0) = \mathcal P_\ell^{(e/o)}(\omega;r_0) D^{(e/o)}(\omega,\epsilon;r_0),
	\label{eq:PerturbedWronskianFactorization}
\end{equation}
where $\mathcal P_\ell^{(e/o)}(\omega;r_0) = \phi_{\omega\ell}^{\mathrm{in},(e/o)}(r_{0*}) \phi_{\omega\ell}^{\mathrm{up},(e/o)}(r_{0*})$ and $D^{(e/o)}=F^{(e/o)}-\epsilon$. At a repelling point satisfying Eq.~\eqref{eq:EPRealityCondition}, the choice $\epsilon=\epsilon_{\rm EP}^{(e/o)}$ gives $D^{(e/o)}=0$, while Eq.~\eqref{eq:RepellingPointCondition} gives $D_\omega^{(e/o)}=0$. It follows immediately that $\mathcal W_\ell^{(e/o)}=0$ and $\partial_\omega\mathcal W_\ell^{(e/o)}=0$ at the critical point.

Differentiating Eq.~\eqref{eq:PerturbedWronskianFactorization} a second time with respect to $\omega$, and evaluating it at the critical point, gives
\begin{equation}
	\left. \frac{\partial^2\mathcal W_\ell^{(e/o)}}{\partial\omega^2} \right|_{\rm EP} = \left. \mathcal P_\ell^{(e/o)} F_{\omega\omega}^{(e/o)} \right|_{\rm EP}.
	\label{eq:EPSecondDerivative}
\end{equation}
Provided that $\mathcal P_\ell^{(e/o)}|_{\rm EP}\neq0$ and $F_{\omega\omega}^{(e/o)}|_{\rm EP}\neq0$, the perturbed Wronskian therefore possesses a zero of algebraic multiplicity two. At the same frequency, the two boundary-condition solutions become linearly dependent and define a single resonance mode. The algebraic multiplicity is therefore two, whereas the geometric multiplicity is one, and the critical point is consequently a second-order EP.

In the following, we label the EPs by $\omega_{\rm EP}^{(n,k,p)}$, $r_{\rm EP}^{(n,k,p)}$, and $\epsilon_{\rm EP}^{(n,k,p)}$, where $n$ identifies the Schwarzschild QNM family, $k$ labels the successive exceptional points along that family, and $p=e,o$ denotes the parity sector. For a given parity, the exceptional-point frequency satisfies
\begin{equation}
	F_\omega^{(p)} \left( \omega_{\rm EP}^{(n,k,p)}; r_{\rm EP}^{(n,k,p)} \right) =0,
	\label{eq:RealEPStationaryCondition}
\end{equation}
where the perturbation location $r_{\rm EP}^{(n,k,p)}$ is selected by
the reality condition
\begin{equation}
	\operatorname{Im} F^{(p)} \left( \omega_{\rm EP}^{(n,k,p)}; r_{\rm EP}^{(n,k,p)} \right) =0.
	\label{eq:RealEPRealityCondition}
\end{equation}
The corresponding critical coupling is then
\begin{equation}
	\epsilon_{\rm EP}^{(n,k,p)} = \operatorname{Re} F^{(p)} \left( \omega_{\rm EP}^{(n,k,p)}; r_{\rm EP}^{(n,k,p)} \right),
	\label{eq:RealEPCriticalCoupling}
\end{equation}
with the second-order character of the exceptional point requiring
\begin{equation}
	F_{\omega\omega}^{(p)} \left( \omega_{\rm EP}^{(n,k,p)}; r_{\rm EP}^{(n,k,p)} \right) \neq0.
	\label{eq:RealEPSecondOrder}
\end{equation}

\section{Chandrasekhar-Detweiler relation and isospectrality breaking}
\label{sec:IsospectralityBreaking}

\subsection{Relation between the normalized Wronskians}
\label{subsec:NormalizedWronskianRelation}

We now use the Chandrasekhar-Detweiler transformation \eqref{eq:CD} to establish an exact relation between the even- and odd-parity normalized Wronskians. Applying Eq.~\eqref{eq:CD} to the boundary-normalized ingoing and upgoing solutions, and using their asymptotic normalizations, we obtain at the position of the defect
\begin{equation}
	\begin{split}
		&\phi_{\omega\ell}^{\mathrm{in},(e)}(r_{0*})
		\phi_{\omega\ell}^{\mathrm{up},(e)}(r_{0*})
		\\
		&=
		\frac{
			\left[
			\left(
			\displaystyle\frac{\dd}{\dd r_*}+\Gamma
			\right)
			\phi_{\omega\ell}^{\mathrm{in},(o)}
			\right]_{r_{0*}}
			\left[
			\left(
			\displaystyle\frac{\dd}{\dd r_*}+\Gamma
			\right)
			\phi_{\omega\ell}^{\mathrm{up},(o)}
			\right]_{r_{0*}}
		}{
			\omega^2+\Omega^2
		}.
	\end{split}
	\label{eq:CDProduct}
\end{equation}
By means of Eqs.~\eqref{eq:WronskianIsospectrality} and
\eqref{eq:NormalizedWronskians}, this gives the exact relation
\begin{equation}
	F^{(e)}(\omega;r_0) = C(\omega;r_0) F^{(o)}(\omega;r_0),
	\label{eq:NormalizedWronskianRelation}
\end{equation}
where
\begin{equation}
	C(\omega;r_0) = \left[ \frac{\omega^2+\Omega^2} { \left( \displaystyle\frac{\dd}{\dd r_*} \ln\phi_{\omega\ell}^{\mathrm{in},(o)} +\Gamma \right) \left( \displaystyle\frac{\dd}{\dd r_*} \ln\phi_{\omega\ell}^{\mathrm{up},(o)} +\Gamma \right) } \right]_{r_{0*}} .
	\label{eq:ChandrasekharFactor}
\end{equation}

Using the definition of the spectral functions in Eq.~\eqref{eq:PerturbedSpectralEquation}, Eq.~\eqref{eq:NormalizedWronskianRelation} then yields
\begin{equation}
	D^{(e)}(\omega,\epsilon;r_0) = C(\omega;r_0) D^{(o)}(\omega,\epsilon;r_0) + \left[ C(\omega;r_0)-1 \right]\epsilon .
	\label{eq:SpectralFunctionRelation}
\end{equation}
This relation is exact for the point-defect model. In particular, if $D^{(o)}(\omega,\epsilon;r_0)=0$, Eq.~\eqref{eq:SpectralFunctionRelation} gives $D^{(e)}(\omega,\epsilon;r_0) = [C(\omega;r_0)-1]\epsilon$. For a nonvanishing defect, the same frequency is therefore a resonance of both parity sectors only if $C(\omega;r_0)=1$. Since this condition is not satisfied in general, an identical localized defect generically breaks the Chandrasekhar-Detweiler isospectral relation.

\subsection{Small-coupling isospectrality breaking}
\label{subsec:SmallCouplingBreaking}

We now consider the behavior of the perturbed frequencies for a weak localized defect. At $\epsilon=0$, the two parity sectors share the same Schwarzschild QNM frequency $\omega_Q^{(n)}$. For sufficiently small $\epsilon$, we write
\begin{equation}
	\omega_n^{(p)} = \omega_Q^{(n)} + \delta\omega_n^{(p)},
	\label{eq:SmallCouplingFrequency}
\end{equation}
with $\delta\omega_n^{(p)}=O(\epsilon)$. Expanding the spectral equation $D^{(p)}(\omega,\epsilon;r_0)=0$ about $\omega_Q^{(n)}$ gives
\begin{equation}
	\delta\omega_n^{(p)} = \frac{\epsilon} {F_\omega^{(p)}(\omega_Q^{(n)};r_0)} + O(\epsilon^2).
	\label{eq:SmallCouplingShift}
\end{equation}

Differentiating Eq.~\eqref{eq:NormalizedWronskianRelation} with respect to $\omega$, and using $F^{(o)}(\omega_Q^{(n)};r_0)=0$, we obtain
\begin{equation}
	F_\omega^{(e)}(\omega_Q^{(n)};r_0) = C(\omega_Q^{(n)};r_0) F_\omega^{(o)}(\omega_Q^{(n)};r_0).
	\label{eq:QNMNormalizedWronskianDerivative}
\end{equation}
Combining Eqs.~\eqref{eq:SmallCouplingShift} and \eqref{eq:QNMNormalizedWronskianDerivative} then gives
\begin{equation}
	\frac{\delta\omega_n^{(e)}} {\delta\omega_n^{(o)}} = \frac{1} {C(\omega_Q^{(n)};r_0)} + O(\epsilon).
	\label{eq:SmallCouplingShiftRatio}
\end{equation}
Thus, although the two parity sectors originate from the same Schwarzschild QNM, an identical localized defect generally produces different complex frequency shifts. At leading order in $\epsilon$, the shifts coincide only when $C(\omega_Q^{(n)};r_0)=1$.

It is also useful to consider the large-distance behavior of the factor $C(\omega;r_0)$ defined in Eq.~\eqref{eq:ChandrasekharFactor}. As $r_{0*}\to+\infty$, one has $\Gamma(r_0)\to\Omega$. Moreover, for the frequencies of interest in the lower half of the complex-frequency plane, the outgoing contribution to $\phi_{\omega\ell}^{\mathrm{in},(o)}$ dominates at large $r_{0*}$, while $\phi_{\omega\ell}^{\mathrm{up},(o)}$ is purely outgoing. The logarithmic derivatives entering Eq.~\eqref{eq:ChandrasekharFactor} therefore both approach $\ii\omega$. It follows that
\begin{equation}
	C(\omega;r_0) \longrightarrow C_\infty(\omega) = \frac{\Omega-\ii\omega} {\Omega+\ii\omega},
	\label{eq:AsymptoticChandrasekharFactor}
\end{equation}
where $C_\infty(\omega)$ denotes the asymptotic Chandrasekhar-Detweiler factor.

Evaluating Eq.~\eqref{eq:AsymptoticChandrasekharFactor} at the Schwarzschild QNM frequency and using Eq.~\eqref{eq:SmallCouplingShiftRatio}, we obtain
\begin{equation}
	\frac{\delta\omega_n^{(e)}} {\delta\omega_n^{(o)}} \longrightarrow \frac{\Omega+\ii\omega_Q^{(n)}} {\Omega-\ii\omega_Q^{(n)}} + O(\epsilon)
	\label{eq:AsymptoticSmallCouplingShiftRatio}
\end{equation}
as $r_{0*}\to+\infty$. Thus, even for a defect situated arbitrarily far from the black hole, the leading small-coupling frequency shifts of the two parity sectors do not in general coincide.

\section{Large-distance parity structure}
\label{sec:LargeDistanceParityStructure}

For the large-distance analysis, it is convenient to introduce the dimensionless tortoise coordinate $\mathcal R=r_{0*}/M$, together with the dimensionless frequency $\widehat{\omega}=2M\omega$. For the asymptotic quantities involving $\mathcal R$ and $\mathcal R^2$, we adopt the same shifted tortoise-coordinate origin as in Ref.~\cite{OuldElHadj:2026vym}, namely
\begin{equation}
	r_*(r) = r+2M\ln\left(\frac{r}{2M}-1\right) -3M+2M\ln2,
	\label{eq:AsymptoticTortoiseCoordinate}
\end{equation}
so that the light ring $r=3M$ is located at $r_*=0$. The Schwarzschild QNM frequencies and the quantity $\Omega$ introduced in Eq.~\eqref{eq:Omega} will accordingly be written as $\widehat{\omega}_Q^{(n)}=2M\omega_Q^{(n)}$ and $\widehat{\Omega}=2M\Omega$.

\subsection{Repelling-point accumulation and parity splitting}
\label{subsec:RepellingPointAsymptotics}

We first consider the large-distance behavior of the RP branches associated with a given Schwarzschild QNM $\omega_Q^{(n)}$. Following the asymptotic analysis developed in Ref.~\cite{OuldElHadj:2026vym}, we assume that $\omega_Q^{(n)}$ is a simple QNM, so that the incoming amplitude $A_\ell^{(-,p)}(\omega)$ has a simple zero at this frequency.

Using Eq.~\eqref{eq:NormalizedWronskians}, together with the large-distance behavior of the ingoing and upgoing solutions and Eq.~\eqref{eq:WronskianAmplitudes}, the normalized Wronskian near $\omega_Q^{(n)}$ takes the asymptotic form
\begin{equation}
	F^{(p)}(\omega;r_0) \simeq \Xi_n^{(p)}(\omega) \left( \omega-\omega_Q^{(n)} \right) e^{-2\ii\omega r_{0*}},
	\label{eq:AsymptoticNormalizedWronskian}
\end{equation}
where $\Xi_n^{(p)}(\omega)$ is independent of $r_{0*}$ at leading
order and varies slowly with $\omega$ compared with the exponential factor.

Differentiating Eq.~\eqref{eq:AsymptoticNormalizedWronskian} and using the repelling-point condition \eqref{eq:RepellingPointCondition} gives
\begin{equation}
	\frac{1} {\omega_r^{(n,p)}-\omega_Q^{(n)}} + \frac{ \left(\Xi_n^{(p)}\right)' }{ \Xi_n^{(p)} } - 2\ii r_{0*} =0,
	\label{eq:AsymptoticRPCondition}
\end{equation}
where the prime denotes differentiation with respect to $\omega$. Expanding for large $r_{0*}$ yields
\begin{equation}
	\omega_r^{(n,p)} - \omega_Q^{(n)} = -\frac{\ii}{2r_{0*}} - \frac{1}{4r_{0*}^2} \left[ \frac{ \left(\Xi_n^{(p)}\right)' }{ \Xi_n^{(p)} } \right]_{\omega_Q^{(n)}} + O(r_{0*}^{-3}).
	\label{eq:RPAsymptoticExpansion}
\end{equation}
The leading term is independent of parity. In terms of the dimensionless variables introduced above, it gives
\begin{equation}
	\mathcal R \left[ \widehat{\omega}_r^{(n,p)} - \widehat{\omega}_Q^{(n)} \right] \longrightarrow -\ii
	\label{eq:DimensionlessRPAccumulation}
\end{equation}
as $\mathcal R\to+\infty$. Thus, both parity sectors reproduce the same universal large-distance accumulation found in Ref.~\cite{OuldElHadj:2026vym}.

The first parity-dependent contribution appears at the next order. From Eqs.~\eqref{eq:NormalizedWronskianRelation} and \eqref{eq:AsymptoticChandrasekharFactor}, the asymptotic coefficients satisfy
\begin{equation}
	\Xi_n^{(e)}(\omega) = C_\infty(\omega) \Xi_n^{(o)}(\omega).
	\label{eq:AsymptoticXiRelation}
\end{equation}
It follows that
\begin{equation}
	\frac{ \left(\Xi_n^{(e)}\right)' }{ \Xi_n^{(e)} } - \frac{ \left(\Xi_n^{(o)}\right)' }{ \Xi_n^{(o)} } = \frac{\dd}{\dd\omega} \ln C_\infty(\omega) = -\frac{2\ii\Omega} {\Omega^2+\omega^2}.
	\label{eq:AsymptoticXiDerivativeRelation}
\end{equation}
Subtracting the odd-parity expansion from the even-parity expansion in Eq.~\eqref{eq:RPAsymptoticExpansion} therefore gives
\begin{equation}
	\omega_r^{(n,e)} - \omega_r^{(n,o)} = \frac{\ii\Omega} { 2\left[ \Omega^2+ \left(\omega_Q^{(n)}\right)^2 \right]r_{0*}^2 } + O(r_{0*}^{-3}).
	\label{eq:RPParitySplitting}
\end{equation}
Equivalently,
\begin{equation}
	\mathcal R^2 \left[ \widehat{\omega}_r^{(n,e)} - \widehat{\omega}_r^{(n,o)} \right] \longrightarrow \frac{ 2\ii\widehat{\Omega} }{ \widehat{\Omega}^2+ \left(\widehat{\omega}_Q^{(n)}\right)^2 }.
	\label{eq:DimensionlessRPParitySplitting}
\end{equation}
Hence the RP branches share the same leading $1/\mathcal R$ accumulation, whereas their even-odd separation first appears at order $1/\mathcal R^2$.

\subsection{Parity-shifted exceptional-point lattices} \label{subsec:ParityShiftedEPLattices}

A real EP is selected along a repelling-point branch by the reality condition \eqref{eq:RealEPRealityCondition}. The leading large-distance behavior of its frequency is therefore already contained in Eq.~\eqref{eq:DimensionlessRPAccumulation}. We now determine the radial distribution of these exceptional points.

Evaluating Eq.~\eqref{eq:AsymptoticNormalizedWronskian} at a distant repelling point and using the leading term of Eq.~\eqref{eq:RPAsymptoticExpansion} gives
\begin{equation}
	F^{(p)} \left( \omega_r^{(n,p)};r_0 \right) \simeq -\frac{\ii} {2e\,r_{0*}} \Xi_n^{(p)} \left( \omega_Q^{(n)} \right) e^{-2\ii\omega_Q^{(n)}r_{0*}}.
	\label{eq:NormalizedWronskianAtRP}
\end{equation}
In terms of $\mathcal R$, the phase of this expression varies asymptotically with $-\operatorname{Re}\widehat{\omega}_Q^{(n)}\mathcal R$. Consequently, the reality condition \eqref{eq:RealEPRealityCondition} implies
\begin{equation}
	\arg\left[ -\ii\, \Xi_n^{(p)} \left( \omega_Q^{(n)} \right) \right] - \operatorname{Re}\widehat{\omega}_Q^{(n)} \mathcal R_{\rm EP}^{(n,k,p)} \simeq k\pi,
	\label{eq:EPPhaseCondition}
\end{equation}
where $\mathcal R_{\rm EP}^{(n,k,p)} =r_*(r_{\rm EP}^{(n,k,p)})/M$.

It follows that consecutive EPs in either parity sector have the same asymptotic spacing,
\begin{equation}
	\mathcal R_{\rm EP}^{(n,k+1,p)} - \mathcal R_{\rm EP}^{(n,k,p)} \longrightarrow \Delta\mathcal R_n = \frac{\pi} {\operatorname{Re}\widehat{\omega}_Q^{(n)}}.
	\label{eq:EPAsymptoticSpacing}
\end{equation}
Thus, the common Schwarzschild QNM spectrum fixes the period of both exceptional-point lattices.

Their relative position follows directly from Eq.~\eqref{eq:AsymptoticXiRelation}. At the Schwarzschild QNM frequency,
\begin{equation}
	\Xi_n^{(e)} \left( \omega_Q^{(n)} \right) = C_\infty \left( \omega_Q^{(n)} \right) \Xi_n^{(o)} \left( \omega_Q^{(n)} \right),
	\label{eq:AsymptoticXiRelationAtQNM}
\end{equation}
so that the phases entering Eq.~\eqref{eq:EPPhaseCondition} differ by $\arg C_\infty(\omega_Q^{(n)})$. Matching the two sequences in $k$ therefore gives
\begin{equation}
	\mathcal R_{\rm EP}^{(n,k,e)} - \mathcal R_{\rm EP}^{(n,k,o)} \longrightarrow \delta\mathcal R_n = \frac{ \arg C_\infty \left( \omega_Q^{(n)} \right) }{ \operatorname{Re}\widehat{\omega}_Q^{(n)} } \pmod{\Delta\mathcal R_n}.
	\label{eq:EPParityOffset}
\end{equation}
Here the two sequences are matched by continuous continuation from their inner members; a relabeling $k\to k+m$ changes the offset only by an integer multiple of the common asymptotic spacing.

The odd- and even-parity EPs therefore form two asymptotically parallel lattices with the same spacing but a finite relative offset determined by the phase of the asymptotic Chandrasekhar-Detweiler factor.

\subsection{Common fragility exponent and parity-dependent prefactors}
\label{subsec:CriticalCouplingAsymptotics}

We now consider the large-distance behavior of the critical coupling. Taking the modulus of Eq.~\eqref{eq:NormalizedWronskianAtRP} gives
\begin{equation}
	\left| F^{(p)} \left( \omega_r^{(n,p)};r_0 \right) \right| \simeq \frac{\chi_n^{(p)}}{\mathcal R} \exp\left[ \operatorname{Im} \widehat{\omega}_Q^{(n)} \mathcal R \right],
	\label{eq:RPWronskianMagnitude}
\end{equation}
where $\chi_n^{(p)}$ is a positive constant independent of $\mathcal R$ at leading order and contains the modulus of $\Xi_n^{(p)}(\omega_Q^{(n)})$ together with the constant dimensional factors.

At a real exceptional point, Eq.~\eqref{eq:RealEPCriticalCoupling} gives $\epsilon_{\rm EP}^{(n,k,p)} = F^{(p)}(\omega_{\rm EP}^{(n,k,p)};r_{\rm EP}^{(n,k,p)})$. The magnitude of the critical coupling therefore obeys
\begin{equation}
	\left| \epsilon_{\rm EP}^{(n,k,p)} \right| \simeq \frac{ \chi_n^{(p)} }{ \mathcal R_{\rm EP}^{(n,k,p)} } \exp\left[ \operatorname{Im} \widehat{\omega}_Q^{(n)} \mathcal R_{\rm EP}^{(n,k,p)} \right].
	\label{eq:CriticalCouplingAsymptotics}
\end{equation}
Since $\operatorname{Im}\widehat{\omega}_Q^{(n)}<0$, the critical coupling decreases exponentially as the exceptional points move to larger distances, with an additional algebraic factor $1/\mathcal R_{\rm EP}^{(n,k,p)}$.

The exponential factor in Eq.~\eqref{eq:CriticalCouplingAsymptotics} is the same in the two parity sectors, since it is fixed entirely by the common Schwarzschild QNM frequency. The parity dependence is contained in the prefactor. From Eq.~\eqref{eq:AsymptoticXiRelation}, evaluated at $\omega_Q^{(n)}$, we obtain
\begin{equation}
	\frac{\chi_n^{(e)}} {\chi_n^{(o)}} = \left| C_\infty \left( \omega_Q^{(n)} \right) \right|.
	\label{eq:CriticalPrefactorRatio}
\end{equation}
Thus, the two parity sectors share the same exponential fragility exponent, while the relative scale of their critical couplings is controlled by the modulus of the asymptotic Chandrasekhar-Detweiler factor.

Together with Eq.~\eqref{eq:EPParityOffset}, this result gives a direct interpretation of the complex quantity $C_\infty(\omega_Q^{(n)})$: its phase determines the relative displacement of the odd- and even-parity exceptional-point lattices, whereas its modulus determines the ratio of their asymptotic critical scales.

\section{Results: parity structure of the perturbed spectrum}
\label{sec:Results}

\subsection{Numerical methods}
\label{subsec:NumericalMethods}

The numerical procedure used for the even-parity sector closely follows that developed for the odd-parity Regge-Wheeler problem in Ref.~\cite{OuldElHadj:2026vym}. We therefore recall only the main ingredients needed for the Zerilli calculation.

For each parity sector, the boundary-normalized homogeneous solutions are obtained by high-precision numerical integration of the Regge-Wheeler or Zerilli equation. The integrations are initialized from series expansions at the event horizon and spatial infinity, with the asymptotic expansions resummed using Padé approximants. An adaptive integration procedure with automatic stiffness switching is used at complex frequencies. The resulting solutions are evaluated at the defect location and used to construct $F^{(p)}(\omega;r_0)$ and its frequency derivatives.

For fixed $r_0$, RPs are obtained from $F_\omega^{(p)}=0$ by a complex Newton-Raphson method and followed by continuation in $r_0$. Real EPs are located from the crossings of the real $F$ axis along these branches and refined with a secant method, while recomputing the RP frequency at each step. The critical coupling is then obtained from Eq.~\eqref{eq:RealEPCriticalCoupling}. For distant EPs, we additionally monitor $|F_\omega^{(p)}|$ and the relative reality residual $|\operatorname{Im}F^{(p)}|/|\operatorname{Re}F^{(p)}|$, and verify that $F_{\omega\omega}^{(p)}\neq0$.

The odd- and even-parity datasets used in the present comparison are constructed with the same tortoise-coordinate convention and numerical accuracy criteria. In particular, the repelling points and exceptional points are computed with $r_*^{(0)}=0$ in Eq.~\eqref{eq:TortoiseCoordinate}. Throughout the numerical calculations, we use the convention $2M=1$. All numerical calculations were performed using \emph{Mathematica}.

\subsection{Small-coupling parity splitting}
\label{subsec:NumericalSmallCouplingSplitting}

Before turning to the exceptional-point structure, we examine the parity splitting in the perturbative regime. Equations \eqref{eq:QNMNormalizedWronskianDerivative} and \eqref{eq:SmallCouplingShiftRatio} show that an identical weak local defect produces different complex frequency shifts in the two parity sectors, with their relative response controlled by the Chandrasekhar-Detweiler factor. The breaking of isospectrality is therefore already present infinitesimally close to the common Schwarzschild spectrum and is not tied to the occurrence of an exceptional point.

For the fundamental $\ell=2$ Schwarzschild QNM, we independently continue the odd- and even-parity resonances from $\epsilon=0$ at fixed $r_0$. Table~\ref{tab:SmallCouplingSplitting} compares the ratio of the resulting frequency shifts with the linear-response value $F_\omega^{(o)}/F_\omega^{(e)}=1/C$, evaluated at the unperturbed QNM. Progressively smaller values of $\epsilon$ are used as the defect is moved outward in order to remain in the linear regime.

\begin{table}[htbp]
	\centering
	\caption{
		Small-coupling parity splitting of the fundamental Schwarzschild QNM for $s=2$ and $\ell=2$. The third column gives the ratio of the even- and odd-parity frequency shifts at the smallest coupling $\epsilon_{\rm min}$ used for each defect location. The fourth column gives the linear-response value
		$F_\omega^{(o)}(\omega_Q^{(0)};r_0)/
		F_\omega^{(e)}(\omega_Q^{(0)};r_0)
		=1/C(\omega_Q^{(0)};r_0)$,
		evaluated at $\epsilon=0$.
	}
	\label{tab:SmallCouplingSplitting}
		\footnotesize
		\setlength{\tabcolsep}{2pt}
	\begin{tabular}{c c c c}
		\hline\hline
		$r_0/M$
		& $\epsilon_{\rm min}$
		& $\delta\omega_0^{(e)}/\delta\omega_0^{(o)}$
		& $1/C(\omega_Q^{(0)};r_0)$
		\\
		\hline
		$10$
		& $10^{-6}$
		& $1.04131547+0.33410212\,\ii$
		& $1.04131107+0.33409884\,\ii$
		\\
		$50$
		& $10^{-10}$
		& $1.01772663+0.39147297\,\ii$
		& $1.01772075+0.39146931\,\ii$
		\\
		$100$
		& $10^{-16}$
		& $1.01647708+0.39350813\,\ii$
		& $1.01647703+0.39350801\,\ii$
		\\
		\hline\hline
	\end{tabular}
\end{table}

At each radius, the frequency-shift ratio approaches the corresponding linear-response value as $\epsilon$ is reduced, confirming that the leading parity splitting is already encoded in the local spectral response of the unperturbed QNM. As $r_0$ increases, the finite-distance factor also approaches $1/C_\infty(\omega_Q^{(0)})=1.01602439+0.39420009\,\ii$, in agreement with Eq.~\eqref{eq:AsymptoticSmallCouplingShiftRatio}.

This perturbative splitting provides a local precursor of the parity-dependent exceptional-point structure studied below. The more detailed role of the asymptotic Chandrasekhar-Detweiler factor in organizing the exceptional-point sequences emerges only in the large-distance analysis.

\subsection{Exceptional points and repelling points}
\label{subsec:NumericalRPEP}

\begin{figure}[htbp]
	\centering
	\includegraphics[width=0.97\linewidth]{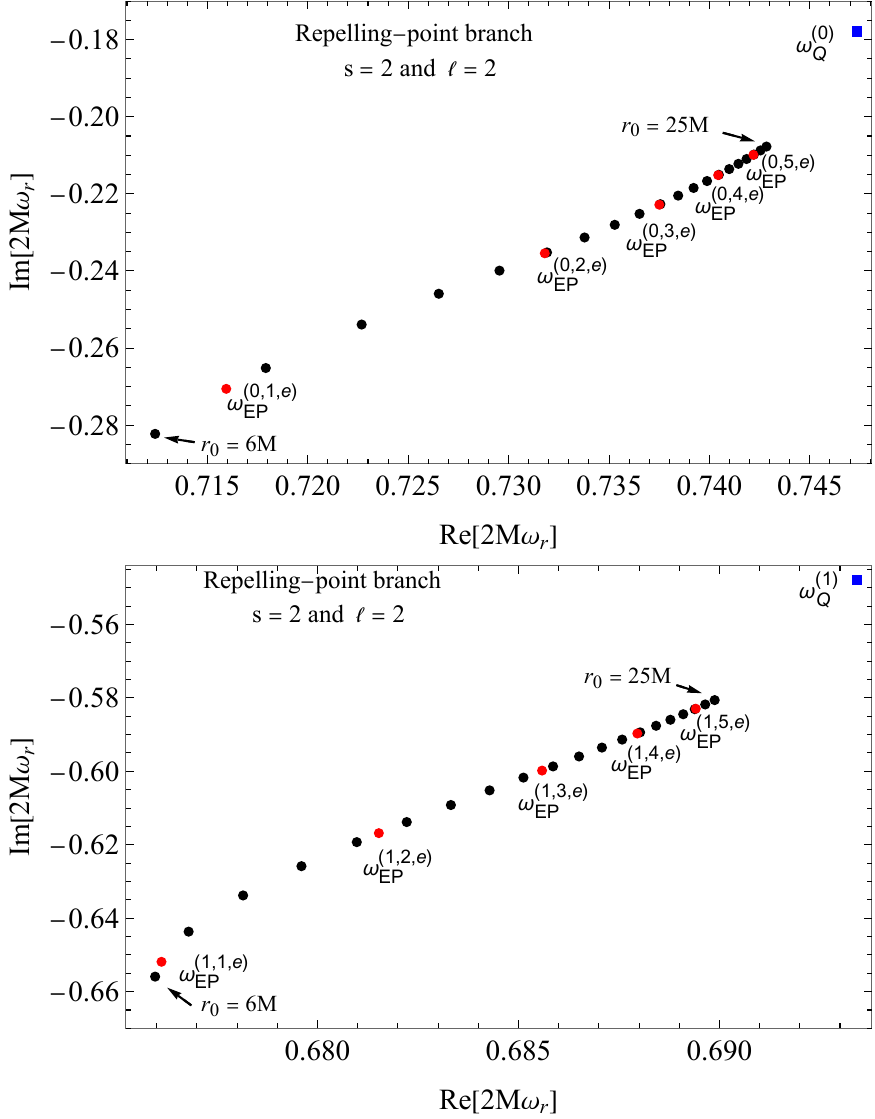}
	\caption{
		Even-parity RP branches associated with the $n=0$ and $n=1$ Schwarzschild QNM families for $s=2$ and $\ell=2$. The black points show the solutions of $F_\omega^{(e)}(\omega_r^{(n,e)};r_0)=0$ as the location $r_0$ of the localized perturbation is varied from $6M$ to $25M$. The blue squares denote the corresponding unperturbed Schwarzschild QNM frequencies $\omega_Q^{(0)}$ and $\omega_Q^{(1)}$. The red points indicate the subset of repelling points for which $\operatorname{Im} F^{(e)}(\omega_r^{(n,e)};r_0)=0$. At these points, the critical perturbation strength $\epsilon_{\rm EP}^{(n,k,e)} = \operatorname{Re} F^{(e)}(\omega_r^{(n,e)};r_0)$ is real, and the corresponding points are exceptional points of the real even-parity perturbed spectral problem. The first five exceptional points of each family are labeled.
	}
	\label{fig:EvenRPBranches}
\end{figure}

We first examine the RP branches in the even-parity sector. Figure~\ref{fig:EvenRPBranches} shows the branches associated with the $n=0$ and $n=1$ Schwarzschild QNM families for $\ell=2$. They are obtained by solving $F_\omega^{(e)}(\omega_r^{(n,e)};r_0)=0$ while varying the defect location $r_0$. The branches accumulate on the corresponding Schwarzschild QNM frequencies as the defect is moved outward, in accordance with the large-distance behavior derived above.

Along each continuous branch, the additional condition $\operatorname{Im}F^{(e)}(\omega_r^{(n,e)};r_0)=0$ selects a discrete sequence of points for which the critical coupling is real. These points, shown in red in Fig.~\ref{fig:EvenRPBranches}, are the even-parity exceptional points of the real perturbed spectral problem. They are labeled by the integer $k$ in order of increasing perturbation radius.

The first five exceptional points of the two families are summarized in Table~\ref{tab:EvenExceptionalPoints}. The table gives their perturbation radii, critical couplings, and complex frequencies. It also makes explicit the radial ordering defining the index $k$. For both families, the sign of the critical coupling alternates along the sequence, while its magnitude decreases rapidly as the defect is moved outward. At the same time, the exceptional frequencies approach the Schwarzschild QNM on which the corresponding repelling-point branch accumulates. The large-distance origin of these features will be examined in the following subsection.

\begin{table}[htbp]
	\caption{
		First five even-parity exceptional points of the $n=0$ and
		$n=1$ families, ordered by increasing perturbation radius
		$r_{\rm EP}^{(n,k,e)}$.
	}
	\label{tab:EvenExceptionalPoints}
	\begin{ruledtabular}
		\begin{tabular}{ccc r@{\,$\times$\,}l c}
			Family & $k$ &
			$r_{\rm EP}^{(n,k,e)}/M$ &
			\multicolumn{2}{c}{$\epsilon_{\rm EP}^{(n,k,e)}$} &
			$2M\omega_{\rm EP}^{(n,k,e)}$ \\
			\hline
			
			$n=0$ & 1 & 6.63294
			& $-7.89667$ & $10^{-2}$
			& $0.715942-0.270576\,\ii$ \\
			
			& 2 & 10.94688
			& $1.61496$ & $10^{-2}$
			& $0.731801-0.235423\,\ii$ \\
			
			& 3 & 14.94413
			& $-5.18868$ & $10^{-3}$
			& $0.737502-0.222821\,\ii$ \\
			
			& 4 & 18.93964
			& $1.85396$ & $10^{-3}$
			& $0.740434-0.215154\,\ii$ \\
			
			& 5 & 22.93214
			& $-7.08446$ & $10^{-4}$
			& $0.742190-0.209892\,\ii$ \\
			
			\hline
			
			$n=1$ & 1 & 6.30681
			& $6.68085$ & $10^{-3}$
			& $0.676119-0.651878\,\ii$ \\
			
			& 2 & 10.42746
			& $-1.28494$ & $10^{-4}$
			& $0.681527-0.616836\,\ii$ \\
			
			& 3 & 14.61324
			& $5.01054$ & $10^{-6}$
			& $0.685588-0.599803\,\ii$ \\
			
			& 4 & 18.83628
			& $-2.56613$ & $10^{-7}$
			& $0.687958-0.589705\,\ii$ \\
			
			& 5 & 23.09299
			& $1.50634$ & $10^{-8}$
			& $0.689412-0.582955\,\ii$ \\
		\end{tabular}
	\end{ruledtabular}
\end{table}

\begin{figure}[htbp]
	\centering
	\includegraphics[width=0.97\linewidth]{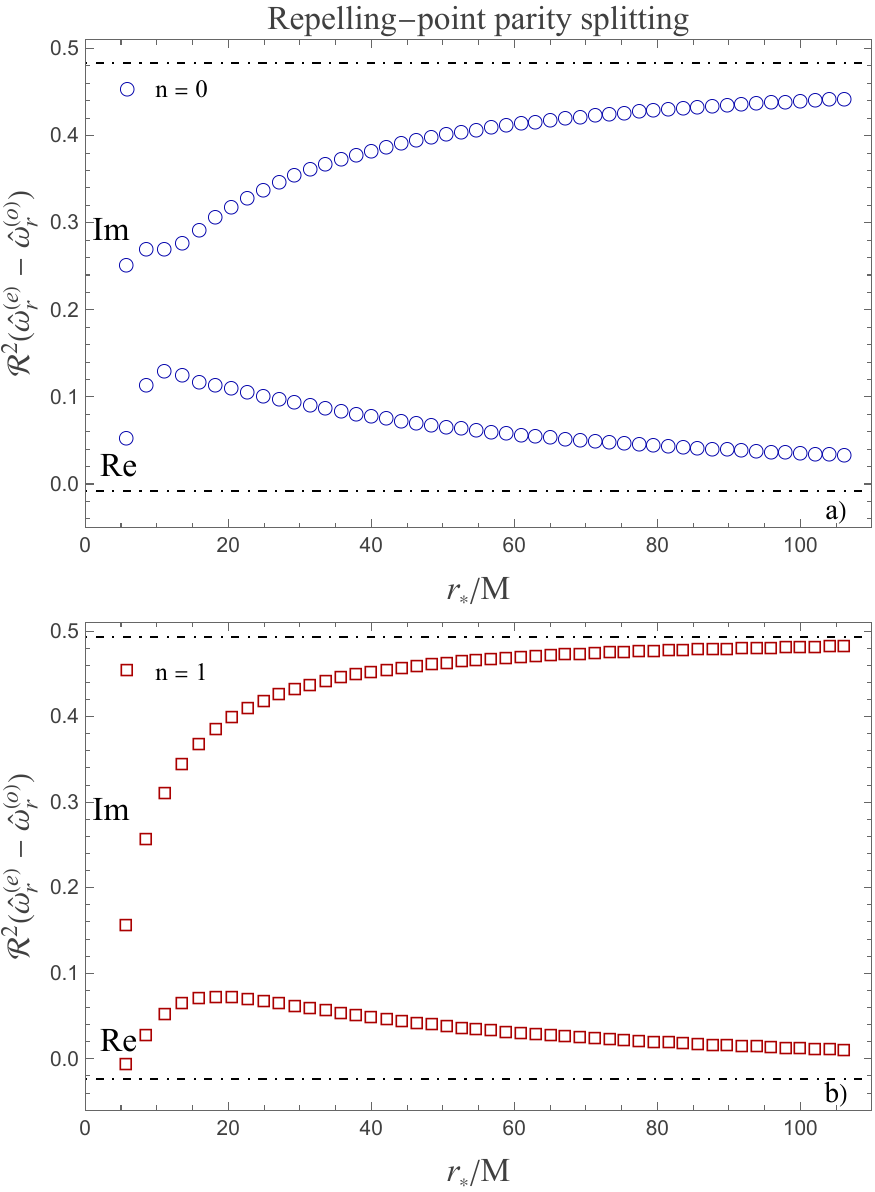}
	\caption{
		Scaled even-odd repelling-point separation for the $n=0$ and $n=1$ Schwarzschild QNM families, shown respectively in panels (a) and (b). The $n=0$ data are represented by blue open circles, while the $n=1$ data are represented by red open squares. In each panel, the real and imaginary parts of $\mathcal R^2 [\widehat{\omega}_r^{(n,e)} -\widehat{\omega}_r^{(n,o)}]$ are shown as the defect location is moved outward, with $\mathcal R=r_{0*}/M$. The upper and lower sequences correspond to the imaginary and real parts, respectively. The horizontal dot-dashed lines show the corresponding asymptotic limits predicted by Eq.~\eqref{eq:DimensionlessRPParitySplitting}, \( 2\ii\widehat{\Omega}/ [\widehat{\Omega}^2+ (\widehat{\omega}_Q^{(n)})^2] \). The convergence of both the real and imaginary parts toward finite limits confirms that the leading even-odd separation of the RP branches occurs at order $1/\mathcal R^2$.
	}
	\label{fig:RPParitySplitting}
\end{figure}

The overall organization closely parallels that found previously for the odd-parity Regge-Wheeler problem~\cite{OuldElHadj:2026vym}. The present comparison, however, allows the parity-dependent displacement of the RP branches to be resolved directly. Since their leading $1/\mathcal R$ accumulation is common to the two sectors, this difference is a subleading effect. To isolate it, Fig.~\ref{fig:RPParitySplitting} shows the real and imaginary parts of $\mathcal R^2 [\widehat{\omega}_r^{(n,e)}-\widehat{\omega}_r^{(n,o)}]$ for the $n=0$ and $n=1$ families and compares them with the parameter-free prediction of Eq.~\eqref{eq:DimensionlessRPParitySplitting}.

For both QNM families, the numerical sequences converge toward the corresponding complex limits predicted by Eq.~\eqref{eq:DimensionlessRPParitySplitting}. This confirms that the leading $1/\mathcal R$ accumulation of the RP branches is parity independent, whereas the breaking of the Chandrasekhar-Detweiler isospectral relation first appears in their relative displacement at order $1/\mathcal R^2$.

\subsection{Large-distance exceptional-point structure}
\label{subsec:NumericalEPAsymptotics}

\begin{figure*}[htbp]
	\centering
	\includegraphics[width=0.85\linewidth]{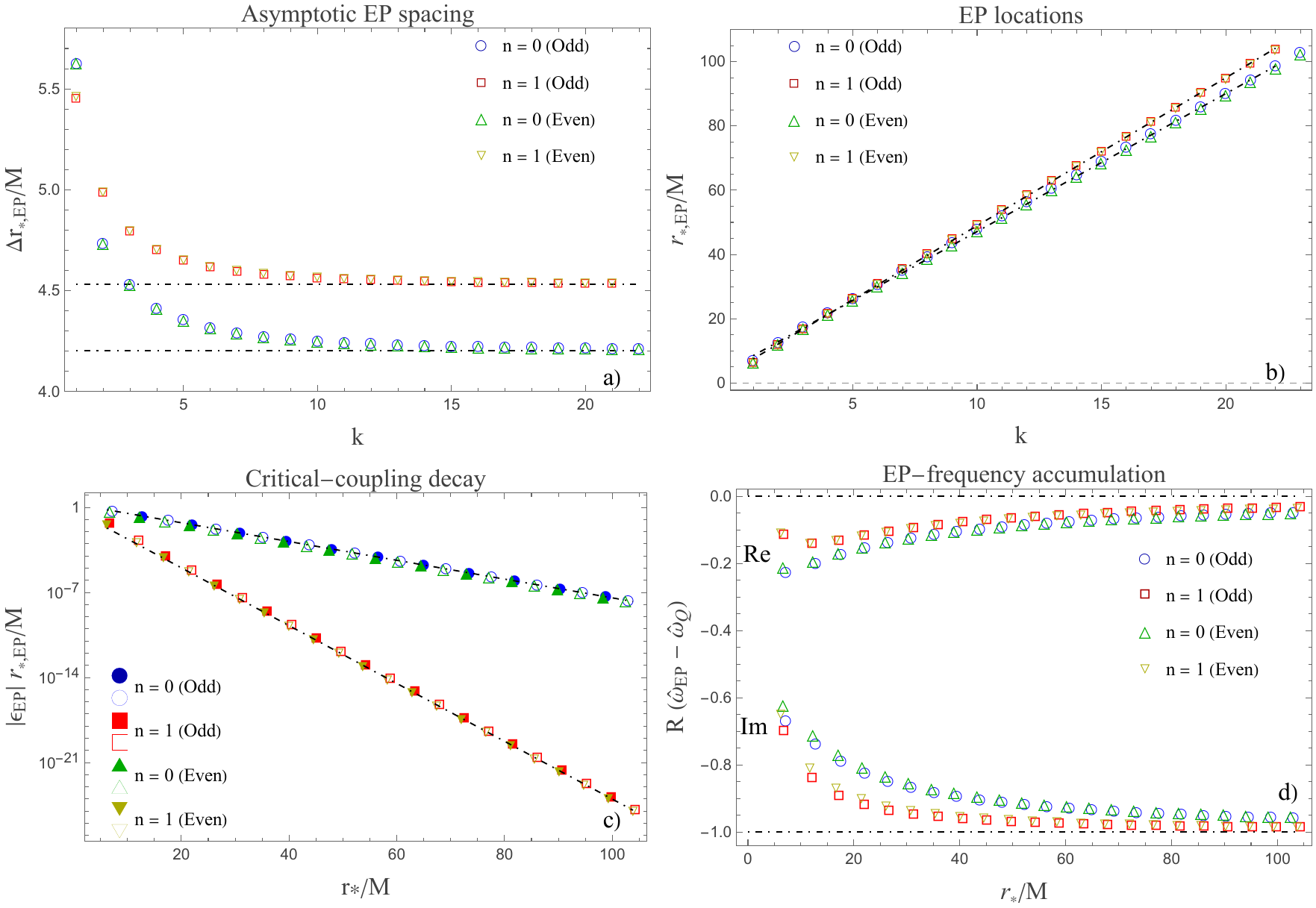}
	\caption{
		Large-distance asymptotics of the odd- and even-parity exceptional-point families associated with the $n=0$ and $n=1$ Schwarzschild QNMs. The four numerical sequences correspond to the $n=0$ and $n=1$ families in the odd- and even-parity sectors, as indicated in the legends, and the dimensionless tortoise-coordinate location $\mathcal R_{\rm EP}^{(n,k,p)}$ is used throughout. (a) Consecutive exceptional-point spacings $\Delta\mathcal R_{\rm EP}^{(n,k,p)} = \mathcal R_{\rm EP}^{(n,k+1,p)} - \mathcal R_{\rm EP}^{(n,k,p)}$. The horizontal dot-dashed lines show the common asymptotic predictions $\pi/\operatorname{Re}\widehat{\omega}_Q^{(n)}$ for each QNM family. The odd- and even-parity sequences associated with the same $n$ approach the same limiting spacing. (b) Exceptional-point locations $\mathcal R_{\rm EP}^{(n,k,p)}$ as functions of the integer label $k$. The dot-dashed lines show the corresponding asymptotic linear behavior, illustrating the emergence of nearly equally spaced exceptional-point lattices at large distance. (c) Corrected critical-coupling magnitudes $|\epsilon_{\rm EP}^{(n,k,p)}| \mathcal R_{\rm EP}^{(n,k,p)}$, displayed on a logarithmic vertical scale. Filled and open symbols correspond respectively to positive and negative critical couplings. The dot-dashed lines have the asymptotic slopes $\operatorname{Im}\widehat{\omega}_Q^{(n)}$, which are common to the two parity sectors for a given QNM family. (d) Real and imaginary parts of $\mathcal R_{\rm EP}^{(n,k,p)} [\widehat{\omega}_{\rm EP}^{(n,k,p)} -\widehat{\omega}_Q^{(n)}]$. The horizontal dot-dashed lines mark the predicted limiting values $0$ and $-1$, respectively. The convergence of all four sequences confirms that the odd- and even-parity exceptional-point families share the same leading large-distance accumulation law, $\mathcal R_{\rm EP}^{(n,k,p)} [\widehat{\omega}_{\rm EP}^{(n,k,p)} -\widehat{\omega}_Q^{(n)}]\rightarrow-\ii$.
	}
	\label{fig:EPAsymptoticsParity}
\end{figure*}

We now examine the large-distance organization of the odd- and even-parity exceptional-point families. Figure~\ref{fig:EPAsymptoticsParity} collects the principal asymptotic diagnostics for the $n=0$ and $n=1$ families in both parity sectors.

For each parity, we define the consecutive exceptional-point spacing by
\begin{equation}
	\Delta\mathcal R_{\rm EP}^{(n,k,p)} = \mathcal R_{\rm EP}^{(n,k+1,p)} - \mathcal R_{\rm EP}^{(n,k,p)}.
	\label{eq:NumericalEPSpacing}
\end{equation}
Panel (a) of Fig.~\ref{fig:EPAsymptoticsParity} shows that, for a given QNM family, the odd- and even-parity sequences approach the same asymptotic spacing predicted by Eq.~\eqref{eq:EPAsymptoticSpacing}. Panel (b) displays the corresponding EP locations and shows that the two parity sequences become nearly parallel at large distance.

Panel (c) shows the corrected critical-coupling magnitudes $|\epsilon_{\rm EP}^{(n,k,p)}| \mathcal R_{\rm EP}^{(n,k,p)}$. For each value of $n$, the odd- and even-parity sequences exhibit the same exponential decay rate, as predicted by Eq.~\eqref{eq:CriticalCouplingAsymptotics}. Finally, panel (d) tests the frequency accumulation through the quantity
\begin{equation}
	\mathcal R_{\rm EP}^{(n,k,p)} \left[ \widehat{\omega}_{\rm EP}^{(n,k,p)} - \widehat{\omega}_Q^{(n)} \right].
	\label{eq:NumericalEPFrequencyAccumulation}
\end{equation}
Its real and imaginary parts approach $0$ and $-1$, respectively, in both parity sectors, in agreement with the universal accumulation law \eqref{eq:DimensionlessRPAccumulation}.

Taken together, the four panels confirm that the two parity sectors share the same leading large-distance structure. They do not imply, however, that the corresponding exceptional-point lattices coincide. The residual parity dependence is revealed by resolving separately their relative positions and critical scales.

\begin{figure}[htbp]
	\centering
	\includegraphics[width=0.97\linewidth]{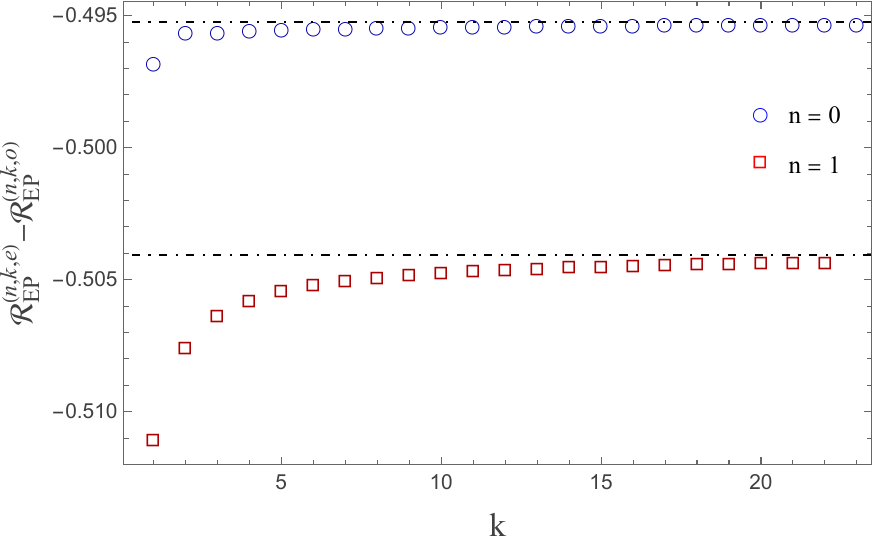}
	\caption{
		Parity-dependent displacement of the exceptional-point lattices associated with the $n=0$ and $n=1$ Schwarzschild QNM families, shown by blue circles and red squares, respectively. For each matched pair of exceptional points carrying the same integer label $k$, the plotted quantity is $\mathcal R_{\rm EP}^{(n,k,e)} -\mathcal R_{\rm EP}^{(n,k,o)}$. The negative values show that the even-parity exceptional points occur at systematically smaller tortoise-coordinate locations than their odd-parity counterparts. The horizontal dot-dashed lines denote the asymptotic offsets $\delta\mathcal R_n$ predicted by Eq.~\eqref{eq:EPParityOffset}, which are determined by the phase of the asymptotic Chandrasekhar-Detweiler factor $C_\infty(\omega_Q^{(n)})$. For both families, the numerical differences converge toward finite nonzero limits as $k$ increases. The two parity sectors therefore form asymptotically parallel exceptional-point lattices with the same spacing but a fixed relative offset.
	}
	\label{fig:EPParityOffset}
\end{figure}

We first consider the relative displacement of the two lattices. Figure~\ref{fig:EPParityOffset} shows $\mathcal R_{\rm EP}^{(n,k,e)} -\mathcal R_{\rm EP}^{(n,k,o)}$ for the $n=0$ and $n=1$ families.

For both QNM families, the numerical differences approach the finite limits predicted by Eq.~\eqref{eq:EPParityOffset}. The odd- and even-parity EP lattices therefore remain distinct as $\mathcal R\to+\infty$: they have the same asymptotic period but are displaced relative to one another by an offset determined by $\arg C_\infty(\omega_Q^{(n)})$.

\begin{figure}[htbp]
	\centering
	\includegraphics[width=0.97\linewidth]{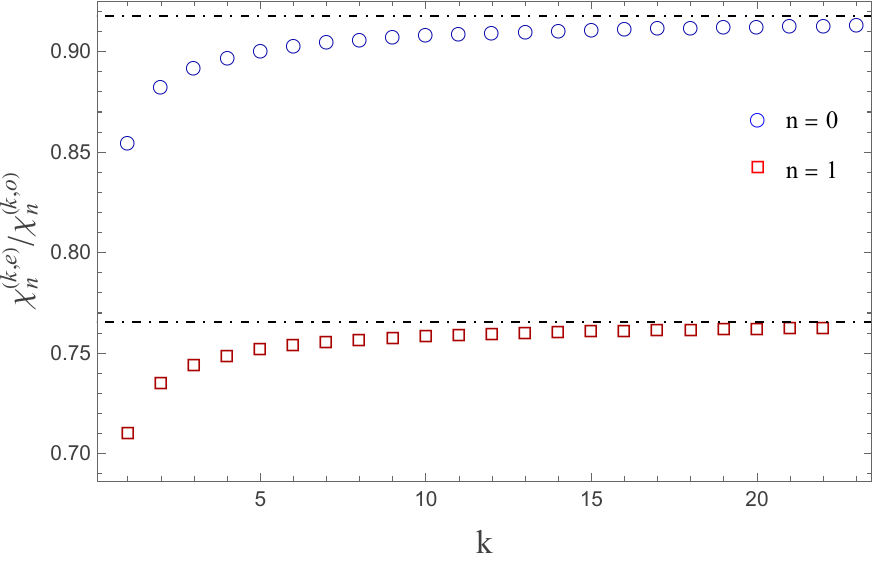}
	\caption{
		Ratio of the finite-distance critical-scale estimators $\chi_{n,k}^{(e)}/\chi_{n,k}^{(o)}$ for the $n=0$ and $n=1$ exceptional-point families, shown by blue circles and red squares, respectively. The quantities $\chi_{n,k}^{(p)}$ are obtained after removing from the critical coupling the common algebraic $1/\mathcal R_{\rm EP}^{(n,k,p)}$ dependence and the exponential decay governed by $\operatorname{Im}\widehat{\omega}_Q^{(n)}$. The horizontal dot-dashed lines show the asymptotic predictions $|C_\infty(\omega_Q^{(n)})|$ given by Eq.~\eqref{eq:CriticalPrefactorRatio}. For both families, the numerical ratios approach finite limits as $k$ increases, confirming that the odd- and even-parity critical couplings share the same exponential fragility exponent while the ratio of their asymptotic prefactors is fixed by the modulus of the Chandrasekhar-Detweiler factor.
	}
	\label{fig:CriticalPrefactorRatio}
\end{figure}

The parity dependence of the critical couplings can be isolated in an analogous manner. For each finite-$k$ exceptional point, we define the critical-scale estimator
\begin{equation}
	\chi_{n,k}^{(p)} = \left| \epsilon_{\rm EP}^{(n,k,p)} \right| \mathcal R_{\rm EP}^{(n,k,p)} \exp\left[ -\operatorname{Im}\widehat{\omega}_Q^{(n)} \mathcal R_{\rm EP}^{(n,k,p)} \right].
	\label{eq:FiniteCriticalPrefactor}
\end{equation}
Equation~\eqref{eq:CriticalCouplingAsymptotics} implies $\chi_{n,k}^{(p)}\to\chi_n^{(p)}$ as $k\to\infty$. Figure~\ref{fig:CriticalPrefactorRatio} therefore compares the ratio $\chi_{n,k}^{(e)}/\chi_{n,k}^{(o)}$ directly with the asymptotic prediction \eqref{eq:CriticalPrefactorRatio}.

The numerical convergence in Fig.~\ref{fig:CriticalPrefactorRatio} confirms that the parity dependence of the critical scale resides in the asymptotic prefactor rather than in the exponential fragility exponent itself. Together, Figs.~\ref{fig:EPParityOffset} and \ref{fig:CriticalPrefactorRatio} provide two complementary tests of the same asymptotic Chandrasekhar-Detweiler factor: its phase determines the relative displacement of the odd- and even-parity exceptional-point lattices, whereas its modulus determines the ratio of their asymptotic critical prefactors.

\subsection{Spectral connectivity and parity contrast}
\label{subsec:NumericalConnectivity}

We finally examine whether the parity dependence of the exceptional points modifies the global connectivity of the resonance spectrum. Following the continuation procedure used in Ref.~\cite{OuldElHadj:2026vym}, we fix the defect location at the relevant exceptional-point value and follow the two coalescing resonance branches back to the Schwarzschild limit $\epsilon=0$. The continuation is performed independently in the two parity sectors, each at its own exceptional-point radius and critical coupling.

\begin{figure}[htbp]
	\centering
	\includegraphics[width=0.97\linewidth]{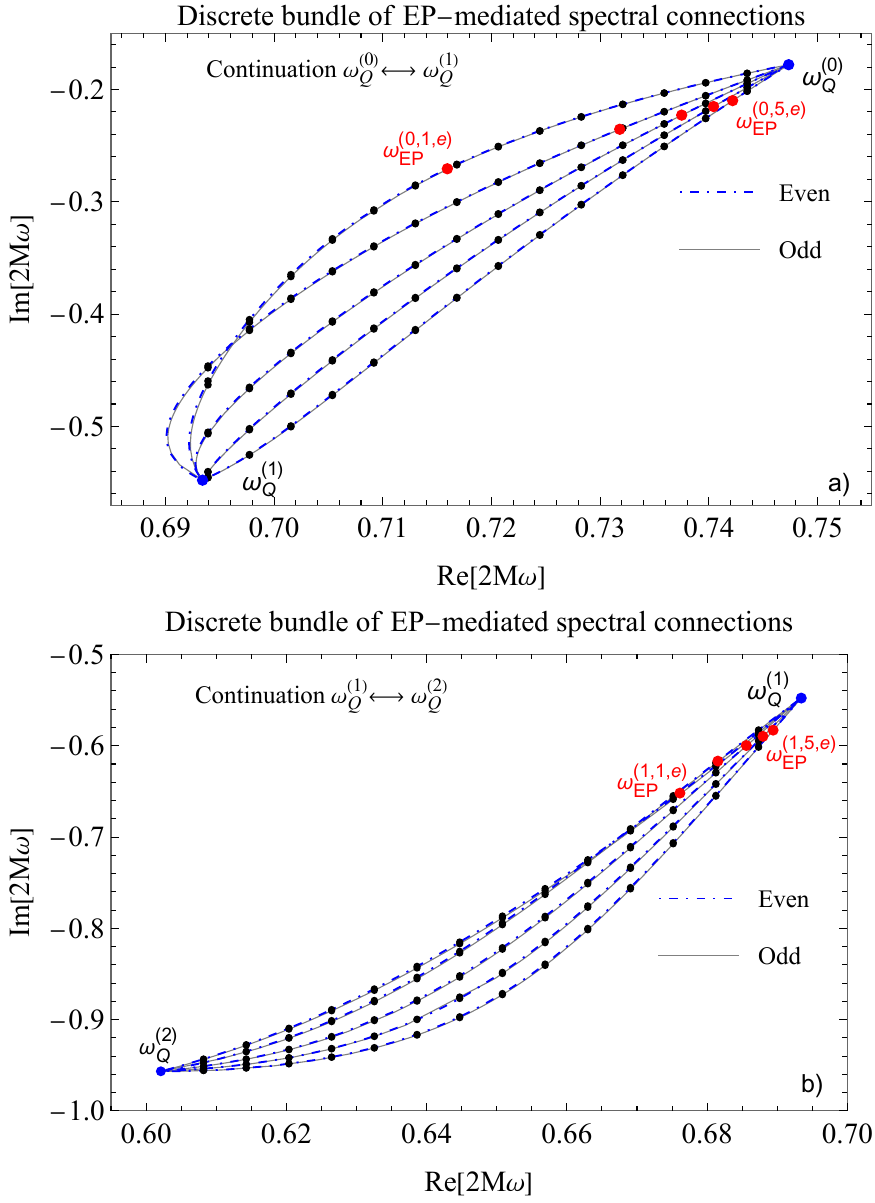}
\caption{
		Discrete bundles of exceptional-point-mediated spectral connections for the first two neighboring pairs of Schwarzschild QNMs, with the even-parity continuations shown by blue dot-dashed curves and the corresponding odd-parity continuations shown in gray for comparison. Panel (a) displays the first five members of the $n=0$ family, each connecting $\omega_Q^{(0)}$ and $\omega_Q^{(1)}$ through a distinct even-parity exceptional point. Panel (b) shows the first five members of the $n=1$ family, connecting $\omega_Q^{(1)}$ and $\omega_Q^{(2)}$. In both panels, the black points indicate the numerical continuation of the even-parity resonance branches and the red points mark the even-parity exceptional frequencies. For each family, the signs of the critical couplings alternate while their magnitudes decrease as the perturbation is moved outward. The gray odd-parity curves provide a direct comparison with the corresponding Regge-Wheeler spectral bridges. Despite the parity dependence of the exceptional-point locations, the odd- and even-parity bundles remain very close in the complex-frequency plane and connect the same neighboring Schwarzschild QNMs.
	}
	\label{fig:EvenEPBundles}
\end{figure}

Figure~\ref{fig:EvenEPBundles} compares the resulting odd- and even-parity spectral connections for the first five exceptional points of the $n=0$ and $n=1$ families. Each member of the $n=0$ family connects $\omega_Q^{(0)}$ to $\omega_Q^{(1)}$, while each member of the $n=1$ family connects $\omega_Q^{(1)}$ to $\omega_Q^{(2)}$.

The continuation therefore establishes, in both parity sectors, the global connectivity
\begin{equation}
	\omega_Q^{(0)} \longleftrightarrow \omega_Q^{(1)} \longleftrightarrow \omega_Q^{(2)}.
	\label{eq:ParityRobustConnectivity}
\end{equation}
This is the same connectivity pattern found previously for the odd-parity Regge-Wheeler problem in Ref.~\cite{OuldElHadj:2026vym}. The remarkable feature of Fig.~\ref{fig:EvenEPBundles} is that the odd- and even-parity bridges remain extremely close in the complex-frequency plane despite being generated at different exceptional-point coordinates.

The critical radii and couplings used in this comparison are collected in Table~\ref{tab:FirstFiveParityEPs} for the first five members of the $n=0$ and $n=1$ families. Each parity sector is evaluated at its own exceptional point; the spectral curves are therefore not being compared at identical values of $(r_0,\epsilon)$.

\begin{table}[htbp]
	\caption{
		First five exceptional points of the $n=0$ and $n=1$ families used in the odd-even spectral-connectivity comparison. Each parity sector is continued at its own critical radius and critical coupling.
	}
	\label{tab:FirstFiveParityEPs}
	\footnotesize
	\setlength{\tabcolsep}{2pt}
	\begin{ruledtabular}
		\begin{tabular}{cccccc}
			Family & $k$
			&
			$r_{\rm EP}^{(n,k,o)}/M$
			&
			$r_{\rm EP}^{(n,k,e)}/M$
			&
			$\epsilon_{\rm EP}^{(n,k,o)}$
			&
			$\epsilon_{\rm EP}^{(n,k,e)}$
			\\
			\hline
			$n=0$ & 1 & 6.98375 & 6.63294
			& $-7.872\times10^{-2}$
			& $-7.896\times10^{-2}$
			\\
			& 2 & 11.35362 & 10.94688
			& $1.610\times10^{-2}$
			& $1.614\times10^{-2}$
			\\
			& 3 & 15.37436 & 14.94413
			& $-5.175\times10^{-3}$
			& $-5.188\times10^{-3}$
			\\
			& 4 & 19.38346 & 18.93964
			& $1.849\times10^{-3}$
			& $1.853\times10^{-3}$
			\\
			& 5 & 23.38486 & 22.93214
			& $-7.067\times10^{-4}$
			& $-7.084\times10^{-4}$
			\\
			\hline
			$n=1$ & 1 & 6.66022 & 6.30681
			& $6.576\times10^{-3}$
			& $6.680\times10^{-3}$
			\\
			& 2 & 10.83962 & 10.42746
			& $-1.269\times10^{-4}$
			& $-1.284\times10^{-4}$
			\\
			& 3 & 15.05139 & 14.61324
			& $4.957\times10^{-6}$
			& $5.010\times10^{-6}$
			\\
			& 4 & 19.28905 & 18.83628
			& $-2.539\times10^{-7}$
			& $-2.566\times10^{-7}$
			\\
			& 5 & 23.55514 & 23.09299
			& $1.491\times10^{-8}$
			& $1.506\times10^{-8}$
		\end{tabular}
	\end{ruledtabular}
\end{table}

To quantify the geometry of these spectral bridges, we follow the arclength construction introduced in Ref.~\cite{OuldElHadj:2026vym}. For the $k$th exceptional point of the $n$th family in parity sector $p$, we introduce the path parameter
\begin{equation}
	\epsilon^{(n,k,p)}(\lambda) = \lambda\epsilon_{\rm EP}^{(n,k,p)}, \qquad 0\leq\lambda\leq1,
	\label{eq:ParityPathParameter}
\end{equation}
and denote by $\widehat{\omega}_j^{(n,k,p)}(\lambda)$, with $j=n,n+1$, the resonance branch connecting the Schwarzschild frequency $\widehat{\omega}_Q^{(j)}$ to $\widehat{\omega}_{\rm EP}^{(n,k,p)}$. Its spectral arclength is defined by
\begin{equation}
	L_j^{(n,k,p)} = \int_0^1 \left| \frac{\dd\widehat{\omega}_j^{(n,k,p)}}{\dd\lambda} \right| \dd\lambda.
	\label{eq:SpectralArclengthParity}
\end{equation}
As in the odd-parity calculation, this definition is independent of the sign of $\epsilon_{\rm EP}^{(n,k,p)}$. Numerically, the integral is evaluated as the polygonal arclength of the ordered continuation data,
\begin{equation}
	L_j^{(n,k,p)} \simeq \sum_{m=1}^{N-1} \left| \widehat{\omega}_{j,m+1}^{(n,k,p)} - \widehat{\omega}_{j,m}^{(n,k,p)} \right|.
	\label{eq:PolygonalSpectralArclengthParity}
\end{equation}

The total length of the spectral bridge is
\begin{equation}
	L_{\rm bridge}^{(n,k,p)} = L_n^{(n,k,p)} + L_{n+1}^{(n,k,p)}.
	\label{eq:TotalBridgeLength}
\end{equation}
The ratio $L_{n+1}^{(n,k,p)}/L_n^{(n,k,p)}$ measures how the total spectral migration is distributed between the two neighboring Schwarzschild branches.

\begin{table}[htbp]
	\caption{
		Arclength geometry of the first five exceptional-point-mediated spectral connections of the $n=0$ and $n=1$ families. For each family, the ratio $L_{n+1}^{(n,k,p)}/L_n^{(n,k,p)}$ measures the relative migration of the higher-overtone branch, while $L_{\rm bridge}^{(n,k,p)}$ is the total length of the two-branch spectral connection.
	}
	\label{tab:BridgeGeometry}
	\footnotesize
	\setlength{\tabcolsep}{2pt}
	\begin{ruledtabular}
		\begin{tabular}{cccccc}
			Family & $k$
			&
			$(L_{n+1}^{(o)}/L_n^{(o)})$
			&
			$(L_{n+1}^{(e)}/L_n^{(e)})$
			&
			$L_{\rm bridge}^{(o)}$
			&
			$L_{\rm bridge}^{(e)}$
			\\
			\hline
			$n=0$ & 1 & 2.84355 & 2.84137 & 0.376833 & 0.376885 \\
			& 2 & 5.31255 & 5.30962 & 0.375839 & 0.375924 \\
			& 3 & 7.15090 & 7.14819 & 0.374503 & 0.374529 \\
			& 4 & 8.87910 & 8.87600 & 0.373947 & 0.373955 \\
			& 5 & 10.54947 & 10.54597 & 0.373863 & 0.373858 \\
			\hline
			$n=1$ & 1 & 2.98315 & 2.97943 & 0.419819 & 0.419742 \\
			& 2 & 5.00978 & 5.00543 & 0.420630 & 0.420561 \\
			& 3 & 7.05020 & 7.04454 & 0.422967 & 0.422827 \\
			& 4 & 9.08900 & 9.09424 & 0.426536 & 0.426287 \\
			& 5 & 11.19382 & 11.18406 & 0.430976 & 0.430748
		\end{tabular}
	\end{ruledtabular}
\end{table}

Table~\ref{tab:BridgeGeometry} shows that the integrated arclength diagnostics are remarkably insensitive to parity in both QNM families. For $n=0$, the ratio $L_1/L_0$ increases from approximately $2.84$ for $k=1$ to $10.55$ for $k=5$, while for $n=1$ the corresponding ratio $L_2/L_1$ increases from approximately $2.98$ to $11.19$. In each case, the odd- and even-parity values remain extremely close. The maximum relative odd-even difference in the total bridge length is approximately $2.3\times10^{-4}$ for the $n=0$ family and $5.9\times10^{-4}$ for the $n=1$ family. The corresponding relative differences in the migration ratios remain below $7.7\times10^{-4}$ and $1.3\times10^{-3}$, respectively.

The two families need not have identical bridge geometries. Indeed, $L_{\rm bridge}$ approaches the Schwarzschild separation between $Q_0$ and $Q_1$ rapidly along the $n=0$ sequence, whereas it increases modestly across the first five members of the $n=1$ family. What is robust here is the odd-even comparison within a given family: at the corresponding exceptional points, the two parity sectors exhibit nearly the same distribution and total amount of spectral migration.

This geometric agreement should not, however, be interpreted as a common exceptional degeneracy at the same control parameters. To make this distinction explicit, we now impose the critical controls of one parity sector on both spectral problems.

This parity selectivity follows directly from Eq.~\eqref{eq:SpectralFunctionRelation}. At an odd-parity exceptional point, $D^{(o)}=D_\omega^{(o)}=0$, whereas, at the same values of $(\omega,r_0,\epsilon)$,
\begin{equation}
	D^{(e)} = (C-1)\epsilon, \qquad D_\omega^{(e)} = C_\omega\epsilon.
	\label{eq:SameControlsParitySelection}
\end{equation}
For a nonvanishing defect, simultaneous exceptional degeneracy of the two sectors would therefore require, at least, the nongeneric conditions $C=1$ and $C_\omega=0$.

\begin{figure}[htbp]
	\centering
	\includegraphics[width=0.97\linewidth]{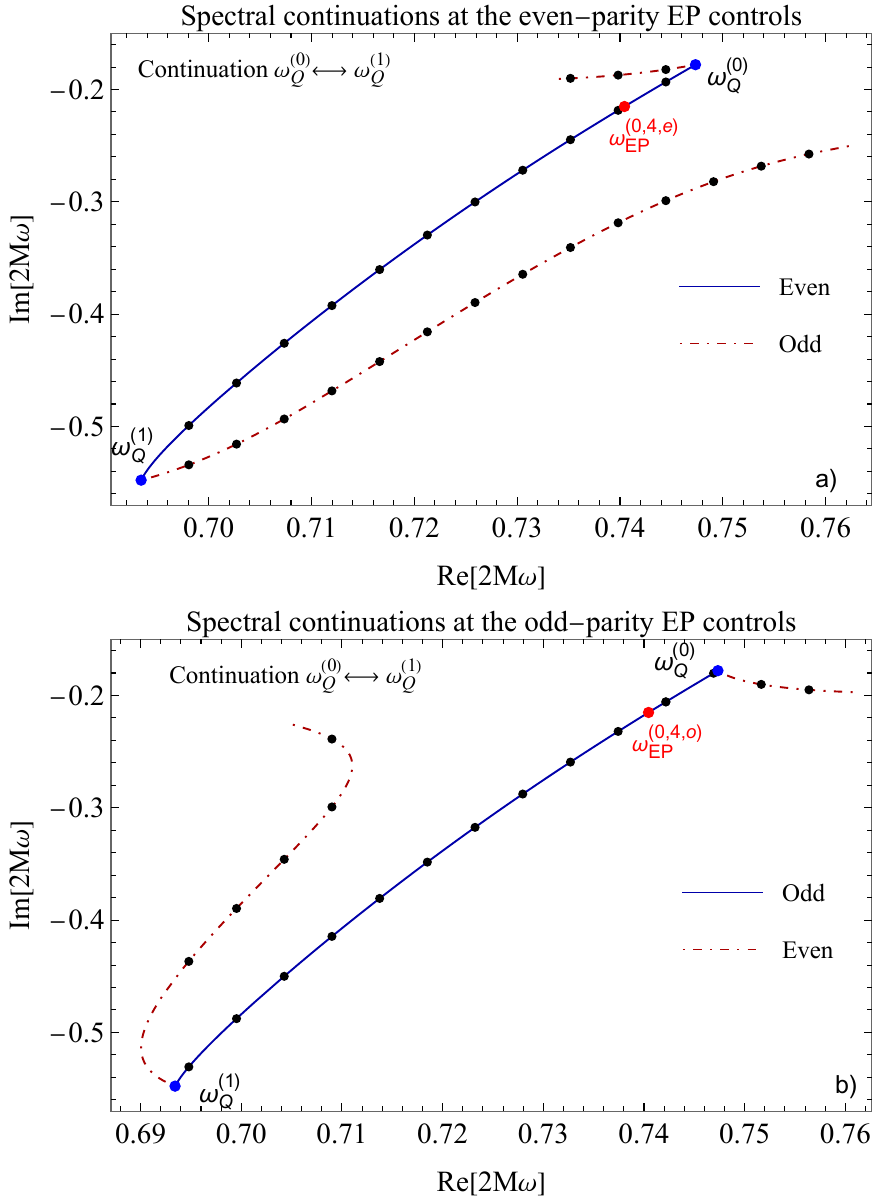}
\caption{
		Spectral continuations of the odd- and even-parity resonance branches under the critical controls of the fourth exceptional point of the $n=0$ family. In both panels, the two parity sectors originate from the same Schwarzschild QNM frequencies $\omega_Q^{(0)}$ and $\omega_Q^{(1)}$ at $\epsilon=0$. Panel (a) uses the even-parity exceptional-point location $r_0=r_{\rm EP}^{(0,4,e)}$ and varies the real coupling through $\epsilon_{\rm EP}^{(0,4,e)}$. The solid blue even-parity branches coalesce at $\omega_{\rm EP}^{(0,4,e)}$, marked by the red point, whereas the red dot-dashed odd-parity continuations remain distinct at the same values of $r_0$ and $\epsilon$. Panel (b) shows the converse situation at $r_0=r_{\rm EP}^{(0,4,o)}$: the solid blue odd-parity branches coalesce at $\omega_{\rm EP}^{(0,4,o)}$, while the red dot-dashed even-parity continuations remain nondegenerate under the same controls. The blue points denote the common unperturbed Schwarzschild QNM frequencies and the black points indicate the numerically continued resonance frequencies. The comparison demonstrates directly that, although the two parity sectors are isospectral in the Schwarzschild limit, the exceptional degeneracy generated by an identical localized defect is parity selective.
	}
	\label{fig:SameControlsContinuations}
\end{figure}

Figure~\ref{fig:SameControlsContinuations} illustrates this directly using the fourth exceptional point of the $n=0$ family. In panel (a), the defect location is fixed at $r_{\rm EP}^{(0,4,e)}$ and the coupling is varied through $\epsilon_{\rm EP}^{(0,4,e)}$. The even-parity branches coalesce at $\omega_{\rm EP}^{(0,4,e)}$, whereas the odd-parity branches remain distinct at the same values of $(r_0,\epsilon)$. Panel (b) shows the converse situation when the odd-parity exceptional-point controls $r_{\rm EP}^{(0,4,o)}$ and $\epsilon_{\rm EP}^{(0,4,o)}$ are imposed.

The parity selectivity of the exceptional degeneracy also has a direct time-domain implication. At the critical controls of parity sector $p$, the coalescence of two simple resonances into an exceptional point produces locally a second-order pole of the corresponding frequency-domain Green function. Its inverse Fourier transform therefore contains the characteristic contribution
\begin{equation}
	\Psi^{(p)}(t) \sim \left(A^{(p)}+B^{(p)}t\right) e^{-\ii\omega_{\rm EP}^{(p)}t}.
	\label{eq:EPRingdown}
\end{equation}
By contrast, under the same values of $(r_0,\epsilon)$, the opposite parity sector remains nondegenerate and its two neighboring resonances remain simple. The corresponding contribution to the ringdown retains the generic form
\begin{equation}
	\Psi^{(\bar p)}(t) \sim A_1^{(\bar p)}e^{-\ii\omega_1^{(\bar p)}t} + A_2^{(\bar p)}e^{-\ii\omega_2^{(\bar p)}t},
	\label{eq:NondegenerateRingdown}
\end{equation}
where $ \omega_1^{(\bar p)}\neq\omega_2^{(\bar p)}$. Thus, the same localized defect can in principle generate a Jordan-type ringdown in one parity sector while leaving the other sector with an ordinary superposition of exponentially damped modes. Equations~\eqref{eq:EPRingdown} and \eqref{eq:NondegenerateRingdown} describe the pole structure implied by the spectral problem; the actual amplitudes depend on the source, initial data, and observable, which are not specified here.

The two comparisons reveal complementary aspects of the parity structure. When each parity sector is continued through its own exceptional point, the global spectral bridges connect the same neighboring Schwarzschild QNMs and closely track one another in the complex-frequency plane, with nearly identical integrated arclength diagnostics. When identical control parameters are imposed on the two sectors, however, the exceptional degeneracy is parity selective, with a correspondingly different pole structure of the associated time-domain response. The breaking of the Chandrasekhar-Detweiler isospectral relation therefore changes the spectral placement of the exceptional points without destroying their global exceptional-point-mediated connectivity.

\section{Discussion and conclusions}
\label{sec:Discussion}

In this article, we have investigated how the exceptional-point structure of the Schwarzschild quasinormal-mode spectrum is modified when the same localized point defect is applied to the odd- and even-parity gravitational master equations. The analysis is a direct continuation of the odd-parity Regge-Wheeler study of Ref.~\cite{OuldElHadj:2026vym}, where localized defects were shown to generate continuous RP branches, discrete families of real EPs, and spectral bridges between neighboring Schwarzschild overtones. The present parity-resolved problem addresses a different question: which elements of this organization are fixed by the common Schwarzschild spectrum, and which reveal the breaking of the Chandrasekhar-Detweiler isospectral relation?

The central analytic result is the exact relation
\[
F^{(e)}(\omega;r_0) = C(\omega;r_0)F^{(o)}(\omega;r_0)
\]
between the two normalized Wronskians. Since an identical defect subtracts the same real quantity $\epsilon$ from both sectors, the corresponding spectral functions satisfy Eq.~\eqref{eq:SpectralFunctionRelation}. Vacuum isospectrality is therefore recovered at $\epsilon=0$, whereas an identical nonzero localized defect generically produces distinct odd- and even-parity resonance frequencies. The same relation determines the leading small-coupling splitting through Eq.~\eqref{eq:SmallCouplingShiftRatio}. The loss of isospectrality is thus not merely a numerical property of the exceptional-point families, but follows directly from the boundary-normalized Chandrasekhar-Detweiler relation.

A particularly simple hierarchy emerges in the large-distance regime. The leading accumulation of the RP branches is parity independent and is governed solely by the common Schwarzschild QNM frequency. The first even-odd separation appears one order later, at $1/\mathcal R^2$. The exceptional points inherit the same hierarchy: the two parity sectors have the same asymptotic lattice spacing, $\pi/\operatorname{Re}\widehat{\omega}_Q^{(n)}$, while the two sequences remain displaced by a finite relative offset determined by the phase of the asymptotic Chandrasekhar-Detweiler factor $C_\infty(\omega_Q^{(n)})$. Likewise, their critical couplings possess the same exponential fragility exponent, fixed by $\operatorname{Im}\widehat{\omega}_Q^{(n)}$, whereas the ratio of their asymptotic prefactors is fixed by $|C_\infty(\omega_Q^{(n)})|$. The phase and modulus of the same complex quantity therefore control two complementary parity effects: the relative placement of the exceptional-point lattices and the relative scale of the perturbation required to reach them.

The exceptional degeneracy itself is parity selective. At the critical radius and coupling of one sector, the other sector is generically not at an exceptional point. This is demonstrated explicitly by the same-control continuations of Fig.~\ref{fig:SameControlsContinuations}: the even-parity branches coalesce when the even-parity critical controls are imposed, while the odd-parity branches remain distinct, and conversely at the odd-parity critical controls. Together with Eq.~\eqref{eq:SameControlsParitySelection}, this shows that the exceptional degeneracy is not shared by the two sectors once the identical defect is switched on.

This parity selectivity also has a direct implication for the local pole contribution to the time-domain response. At an exceptional point of parity sector $p$, the coalescence of two simple resonances produces a second-order pole of the corresponding frequency-domain Green function. For generic excitation and observation, its inverse transform therefore contains the characteristic contribution $(A^{(p)}+B^{(p)}t)e^{-\ii\omega_{\rm EP}^{(p)}t}$. Under the same values of $(r_0,\epsilon)$, the opposite parity sector remains nondegenerate, with two distinct simple resonances, and the corresponding contribution retains the generic form $A_1^{(\bar p)}e^{-\ii\omega_1^{(\bar p)}t} +A_2^{(\bar p)}e^{-\ii\omega_2^{(\bar p)}t}$. Thus the same localized defect can distinguish the two parity sectors not only through their resonance frequencies but also through the analytic structure governing their ringdown contributions. The actual amplitudes, however, depend on the source, initial data, and observable and are not determined by the spectral calculation alone.

The global spectral organization is nevertheless considerably more robust than the exceptional-point coordinates. Continuation through the first five exceptional points of the $n=0$ and $n=1$ families shows that both parity sectors generate the same sequence
\[
\omega_Q^{(0)} \longleftrightarrow \omega_{\rm EP}^{(0,k,p)} \longleftrightarrow \omega_Q^{(1)} \longleftrightarrow \omega_{\rm EP}^{(1,k,p)} \longleftrightarrow \omega_Q^{(2)},
\]
The arclength diagnostics show that this robustness extends beyond the identity of the Schwarzschild endpoints. In both families, the odd- and even-parity total bridge lengths and migration ratios agree at the $10^{-3}$ level or better over the five connections considered. Direct comparison of the corresponding continuations further shows that the odd- and even-parity bridges closely track one another in the original complex-frequency plane, without translation, rescaling, or alignment.

These results motivate a useful distinction between \emph{spectral placement} and \emph{spectral architecture}. The exceptional-point coordinates in control-parameter and frequency space are sensitive to parity, as are the critical couplings required to reach them. By contrast, the neighboring-mode connectivity and the integrated arclength characteristics of the associated spectral bridges are remarkably insensitive to parity. In this sense, breaking the Chandrasekhar-Detweiler isospectral relation does not imply a breaking of the exceptional-point-mediated connectivity of the spectrum.

The boundary-normalized Chandrasekhar-Detweiler relation provides a particularly transparent interpretation of this structure. The local factor $C(\omega;r_0)$ determines how the two normalized spectral functions depart from one another under an identical defect, while its large-distance limit $C_\infty(\omega)$ controls the asymptotic parity offset and relative critical scale. The present model therefore provides a simple example in which two master problems that are exactly isospectral in vacuum acquire distinct resonance spectra under identical local deformations, while retaining much of their common global spectral organization.

In the language of \emph{The Princess and the Pea}, Ref.~\cite{OuldElHadj:2026vym}, replacing the Regge-Wheeler ``mattress'' by its Zerilli counterpart does not make the Princess any less sensitive to the same ``pea'': exponentially weak localized defects again generate RP branches and exceptional-point families with the same leading asymptotic organization and fragility hierarchy. The new element is that the two parity sectors do not feel the same ``pea'' in exactly the same spectral way. Their exceptional points are displaced relative to one another, and an identical defect can place one sector at an exceptional degeneracy while leaving the other nondegenerate.

Several extensions follow naturally. Finite-width perturbations provide an immediate setting in which to test whether the parity-dependent lattice displacement and the parity robustness of the spectral bridges persist beyond the point-defect limit. A particularly interesting next step would be to embed the localized perturbation in a covariantly specified physical model, for example through an environmental stress-energy distribution or a modification of the underlying field equations. Deriving the corresponding odd- and even-parity master equations would make it possible to determine how the parity-dependent frequency shifts and pole structure identified here propagate into mode excitation and observable gravitational-wave signals ~\cite{Silva:2026jih,Bah:2026aia}. This would also provide a natural setting in which to compare frequency isospectrality, resonant pole structure, excitation amplitudes, and the resulting time-domain waveforms. Extending the analysis to rotating black holes would provide a further connection with the exceptional-point and avoided-crossing structures already known in the Kerr spectrum.

\begin{acknowledgments}
	I thank Sam R. Dolan for discussions and for our collaboration on the preceding work that motivated the present study.
\end{acknowledgments}

\bibliography{refs_exceptional_points_even_odd_isospectrality}

\end{document}